\pdfoutput=1

\documentclass[longauth]{aa}  

\usepackage{graphicx}
\usepackage{unicode}
\usepackage{graphicx}
\usepackage{txfonts}
\begin{document}

   \title{THOR - The \ion{H}{i}, OH, Recombination Line Survey of the Milky Way}

   \subtitle{The pilot study - \ion{H}{i} observations of the giant molecular cloud W43}

      \author{
      			S. Bihr\inst{1}
          \and
          		H. Beuther\inst{1}
          \and
          		J. Ott\inst{2}          
          \and
          		K.G. Johnston\inst{3}           
 		  \and
		  		A. Brunthaler\inst{4}    
		  \and
		  		L. D. Anderson\inst{5}
          \and
		  		F. Bigiel\inst{6}
		  \and
		  		P. Carlhoff\inst{7}
		  \and
		  		E. Churchwell\inst{8}
		  \and		  
		        S.C.O. Glover\inst{6}
          \and
          		P.F. Goldsmith\inst{9}
		  \and
		  		F. Heitsch\inst{10}
          \and
          		T. Henning\inst{1}
		  \and
		  		M.H. Heyer\inst{11}         		
		  \and
		  		T. Hill\inst{12}          		
 		  \and
		  		A. Hughes\inst{1}
		  \and
  				R.S. Klessen\inst{6,13,14}
		  \and
		  		H. Linz\inst{1}
		  \and
		  		S.N. Longmore\inst{15}
		  \and
          		N.M. McClure-Griffiths\inst{16}		  		
		  \and
		  		K.M. Menten\inst{4}  
		  \and
				F. Motte\inst{17}  
		  \and
		  		Q. Nguyen-Lu’o’ng\inst{18}
		  \and
		  		R. Plume\inst{19}		  		
		  \and
		  		S.E. Ragan\inst{1}
		  \and
		  		N. Roy\inst{4,20}		  		
		  \and
		 		P. Schilke\inst{7}
		  \and
		  		N. Schneider\inst{21}
		  \and
		  		R.J. Smith\inst{6}
		  \and		
		  		J.M. Stil\inst{19}
		  \and
		 		 J.S. Urquhart\inst{4}
		  \and
		 		A.J. Walsh\inst{22}
		  \and
		 		F. Walter\inst{1}				 		
          }

    \institute	{		Max Planck Institute for Astronomy, K\"onigstuhl 17, 69117 Heidelberg, Germany\\
	  					\email{name@mpia.de}
    			\and
    					National Radio Astronomy Observatory, P.O. Box O, 1003 Lopezville Road, Socorro, NM 87801, USA
         		\and
         				School of Physics and Astronomy, University of Leeds, Leeds LS2 9JT, UK
         		\and
         				Max-Planck-Institut f\"ur Radioastronomie, Auf dem H\"ugel 69, D-53121 Bonn, Germany   
          		\and
         				Department of Physics and Astronomy, West Virginia University, Morgantown, WV 26506, USA
       			\and
    					Universit\"at Heidelberg, Zentrum f\"ur Astronomie, Institut f\"ur Theoretische Astrophysik, Albert-Ueberle-Str. 2, D-69120 Heidelberg, Germany
         		\and
         				1. Physikalisches Institut, Universit\"at zu K\"oln, Z\"ulpicher Str. 77, 50937 K\"oln, Germany         		
          		\and
         				Department of Astronomy, University of Wisconsin, Madison, WI 53706, USA        		
         		\and
         				Jet Propulsion Laboratory, California Institute of Technology, 4800 Oak Grove Drive, Pasadena, CA 91109, USA
         		\and
         				Department of Physics and Astronomy, University of North Carolina-Chapel Hill, Chapel Hill, NC 27599-3255, USA      		
         		\and
         				Department of Astronomy, University of Massachusetts, Amherst, MA 01003-9305, USA      		
           		\and
         				Joint ALMA Observatory, Alonso de Cordova 3107, Vitacura 763-0355, Santiago, Chile     		
      			\and
      					Kavli Institute for Particle Astrophysics and Cosmology, Stanford University, SLAC National Accelerator Laboratory, Menlo Park, CA 94025, USA
      			\and
      					Department of Astronomy and Astrophysics, University of California, 1156 High Street, Santa Cruz, CA 95064, USA 
          		\and
         				Astrophysics Research Institute, Liverpool John Moores University, 146 Brownlow Hill, Liverpool L3 5RF, UK     		
          		\and
         				Australia Telescope National Facility, CSIRO Astronomy and Space Science, Marsfield, NSW 2122, Australia
         		\and
         				Laboratoire AIM, CEA/DSM-CNRS-Universit\'e Paris Diderot, IRFU/Service dAstrophysique, Saclay, 91191 Gif-sur-Yvette, France         		
           		\and
         				Canadian Institute for Theoretical Astrophysics, University of Toronto, 60 St. George Street, Toronto, ON M5S 3H8, Canada    		
         		\and
         				Department of Physics and Astronomy, University of Calgary, 2500 University Drive NW, Calgary, AB T2N 1N4, Canada.
				\and         		
         				Department of Physics, Indian Institute of Technology Kharagpur, Kharagpur 721302, India
         		\and
         				Universit\'e de Bordeaux, Laboratoire d’Astrophysique de Bordeaux, CNRS/INSU, 33270 Floirac, France
           		\and
         				International Centre for Radio Astronomy Research, Curtin University, GPO Box U1987, Perth WA 6845, Australia
					}
          
   \date{Received November 20, 2014; accepted May 11, 2015}

  \abstract{To study the atomic, molecular and ionized emission of Giant Molecular Clouds (GMCs) in the Milky Way, we have initiated a Large Program with the Karl G. Jansky Very Large Array (VLA): `THOR - The \ion{H}{i}, OH, Recombination Line survey of the Milky Way'. We map the 21cm \ion{H}{i} line, 4 OH lines, up to 19 H$\alpha$ recombination lines and the continuum from 1 to 2\,GHz of a significant fraction of the Milky Way ($l=15\degr-67\degr$, $|b| \leq 1\degr$) at an angular resolution of $\sim20\arcsec$. Starting in 2012, we mapped 4 square degrees of the GMC associated with the W43 star-formation complex as a pilot study. The rest of the THOR survey area was observed during 2013 and 2014. In this paper, we focus on the \ion{H}{i} emission from the W43 GMC complex. Classically, the \ion{H}{i} 21cm line is treated as optically thin with properties such as the column density calculated under this assumption. This approach might give reasonable results for regions of low-mass star-formation, however, it is not sufficient to describe giant molecular clouds. We analyzed strong continuum sources to measure the optical depth along the line of sight, and thus correct the \ion{H}{i} 21cm emission for optical depth effects and weak diffuse continuum emission. Hence, we are able to measure the \ion{H}{i} mass of this region more accurately and our analysis reveals a lower limit for the \ion{H}{i} mass of $M = 6.6_{-1.8}\times 10^6$\,M$_{\odot}$ ($v_{\rm{LSR}} = 60-120$\,km\,s$^{-1}$), which is a factor of 2.4 larger than the mass estimated with the assumption of optically thin emission. The \ion{H}{i} column densities are as high as $N_{\rm{\ion{H}{i}}}\sim 150$\,M$_{\odot}$\,pc$^{-2} \approx 1.9 \times 10^{22}\,\rm{cm^{-2}}$, which is an order of magnitude higher than for low mass star formation regions. This result challenges theoretical models that predict a threshold for the \ion{H}{i} column density of $\sim 10$\,M$_{\odot}$\,pc$^{-2}$, at which the formation of molecular hydrogen should set in. By assuming an elliptical layered structure for W43, we estimate the particle density profile. For the atomic gas particle density, we find a linear decrease toward the center of W43 with values decreasing from $n_{\rm{\ion{H}{i}}} = 20$\,cm$^{-3}$ near the cloud edge to almost 0\,cm$^{-3}$ at its center. On the other hand, the molecular hydrogen, traced via dust observations with the Herschel Space Observatory, shows an exponential increase toward the center with densities increasing to $n_{\rm{H_2}} > 200$\,cm$^{-3}$, averaged over a region of $\sim$10\,pc. While at the cloud edge atomic and molecular hydrogen are well mixed, the center of the cloud is strongly dominated by H$_{\rm{2}}$ emission. We do not identify a sharp transition between hydrogen in atomic and molecular form. Our results are an important characterization of the atomic to molecular hydrogen transition in an extreme environment and challenges current theoretical models.}


   \maketitle
\section{Introduction}
It is well known that stars form in Giant Molecular Clouds \citep[GMCs,][]{MacLow2004, McKee2007, Dobbs2014, Offner2014}, which primarily consist of molecular hydrogen. Yet it is still under debate whether molecular hydrogen is actually necessary for star formation or whether molecular hydrogen and stars form only under the same conditions side by side \citep{Glover2012}. The density within these clouds is high enough (particle density: $n>1-100$\,cm$^{-3}$, column density: $N>1-100$\,M$_{\odot}$\,pc$^{-2} \sim 10^{20}-10^{22}$\,cm$^{-2}$) for molecular hydrogen to become self-shielded from the interstellar radiation field, which would otherwise dissociate the H$_{\rm{2}}$ molecules \citep[e.g.][]{Dobbs2014}. Hence, molecular clouds form in the interior of large clouds of neutral hydrogen, which themselves are the environment of molecular clouds. Another open question is the fraction of the neutral hydrogen within molecular clouds and how this affects the physical conditions \citep{Krco2010, Goldsmith2005}. Furthermore, the correlation between the transition between these two fundamental states of hydrogen and the corresponding physical conditions are also still unclear.\\
As cold molecular hydrogen is challenging to observe directly, it is difficult to study its distribution in detail. Classically, the low-J rotational transitions of CO are used as a tracer for H$_{\rm{2}}$. However, recent simulations and observations show, that a large amount ($\sim 40 \%$) of H$_{\rm{2}}$ is CO-dark and therefore not well traced by CO \citep{Pineda2013, Smith2014}. Another approach to study molecular hydrogen is via observations of thermal dust emission or dust extinction \citep{Lada2007,Molinari2010,Kainulainen2013,Kainulainen2013b}. On the other hand, the 21cm spin-flip transition of hydrogen offers a well-known method to measure the atomic gas content. Even though the 21cm line is well studied \citep[e.g.][]{Radhakrishnan1972, Gibson2000, Taylor2003, Gibson2005, Heiles2003a, Heiles2003b, Strasser2004, Goldsmith2005, Stil2006, Kalberla2009, McClure-Griffiths2012, Roy2013, Fukui2015, Murray2014, Motte2014}, it is difficult to disentangle the different contributions of the cold and warm \ion{H}{i} in emission and absorption for different spin temperatures and optical depths. On top of that, radio continuum emission at the frequency of the \ion{H}{i} emission line can also suppress the intensity of the observed \ion{H}{i} emission, an effect that is especially significant for the Galactic Plane.\\
To address cloud formation, \ion{H}{i} to H$_{\rm{2}}$ formation as well as many other issues in ISM studies of the Milky Way, we initiated the THOR project: The \ion{H}{i}, OH, Recombination Line Survey of the Milky Way (Beuther et al. in prep.). We are using the Karl G. Jansky Very Large Array (VLA) in a large program to observe the \ion{H}{i} line at 21cm, four OH lines at 1612, 1665, 1667 and 1720MHz, 19 H$\alpha$ radio recombination lines (RRL) and the continuum from $1-2$\,GHz, which is split in 8 sub bands. These observations result in an angular resolution of $15-20\arcsec$. We have been awarded more than 200 hours observing time to map a significant fraction of the Milky Way (Galactic longitude $l = 15 - 67$\degr, Galactic latitude $|b| \leq 1$\degr) for the semester 2013A and 2014B. To test our observing strategy and data reduction, we began with a pilot study of a $2\degr\times2\degr$ field around W43, which was completed in 10 hours. The following paper describes this pilot study, which was observed in the semester 2012A. As the wealth of this data set is immense, we will focus on the \ion{H}{i} line in this paper, but consecutive papers will focus on other aspects of the pilot study (e.g. Walsh et al. in prep., Johnston et al. in prep.). The full survey will be presented in Beuther et al. (in prep.).\\
The field chosen for the pilot study is around the massive star forming complex W43 ($l = 29.2-31.5$\degr, $|b| \leq 1$\degr). This region is situated at the intersection of the Galactic bar and the first spiral arm \citep[Scutum-Centaurus Galactic arm,][]{Nguyen2011,Carlhoff2013}, leading to complex kinematic structures and possibly high star formation activity. W43 is referred to as a Galactic mini-starburst region \citep{Motte2003, Bally2010} and shows a star-formation rate of $\sim 0.1$\,M$_{\odot}$\,yr$^{-1}$ \citep{Nguyen2011} or 5-10\% of the star formation rate in the entire Milky Way. \citet{Motte2014} found velocity gradients in CO and \ion{H}{i} 21cm emission. These velocity gradients could indicate large scale velocity flows, which could cause the vast star-formation activity. \citet{Motte2014} also discussed that they do not find a threshold for the \ion{H}{i} column density, which is proposed by theoretical models \citep{Krumholz2008,Krumholz2009}. They argue, that we see several transition layers of \ion{H}{i} and H$_{\rm{2}}$ along the line of sight and that such models are not suited to describe complicated molecular cloud complexes such as W43. The center of W43 harbors a large \ion{H}{ii} region which is fueled by a Wolf-Rayet and OB star cluster \citep{Liszt1993,Lester1985,Blum1999}. Furthermore, W43 exhibits several high-mass starless molecular clumps, which are still in an early stage of star-formation \citep{Beuther2012, Louvet2014}. Some massive dense clumps can potentially form Young Massive Clusters, progenitors of globular clusters \citep{Louvet2014,Nguyen2013}. The BeSSeL survey \citep[Bar and Spiral Structure Legacy Survey,][]{Brunthaler2011, Reid2014} has determined the distance to W43 to be 5.5$\pm$0.4\,kpc from parallax measurements of methanol and water masers \citep{Zhang2014}. This result has to be treated cautiously, as none of the four masers used for the parallax measurements are spatially directly associated with W43-Main.\\
\section{Observations and data reduction}
\subsection{VLA Observations}
\begin{figure}
  \resizebox{\hsize}{!}{\includegraphics{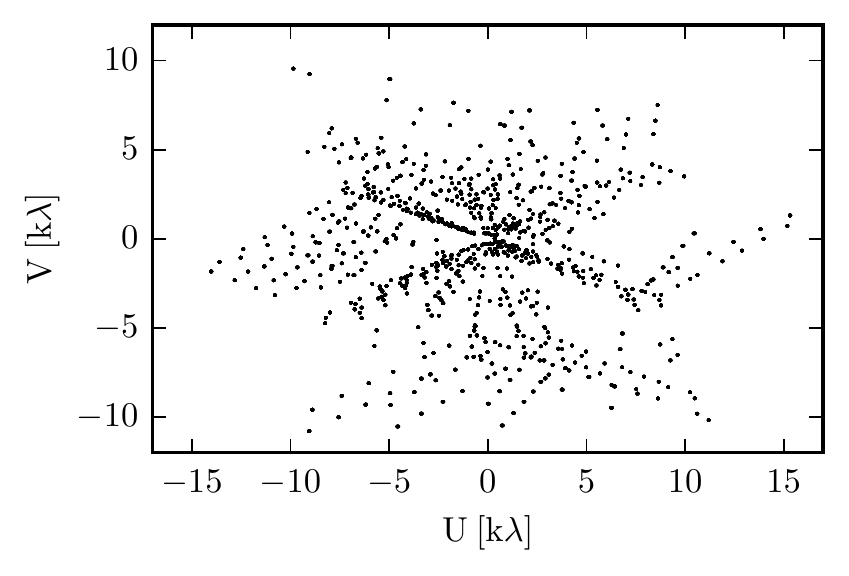}}
  \caption{UV coverage of one pointing (centered at RA 18:47:54.125746, Dec -03:28:42.90941, J2000) after 4 $\times$ 2\,min observing time.}
  \label{fig_uv_coverage}
\end{figure}

We mapped a $2\degr\times2\degr$ field around W43 ($l = 29.2-31.5$\degr, $|b| \leq 1$\degr) during the 2012A semester with the Karl G. Jansky Very Large Array (VLA) in New Mexico in C-configuration (Project 12A-161). As we used Nyquist sampling at 1.42\,GHz with a primary beam size of 32\arcmin, we needed 59 pointings to cover the 4 square degree mosaic. We chose a hexagonal geometry for the mosaic, similar to the VGPS survey \citep{Stil2006}. This results in a smooth areal sensitivity function with fluctuations of less than 1\% in the inner region and a decreasing sensitivity toward the edges of our field. Each pointing was observed 4 $\times$ 2\,min, which results in an overall observation time of 10 hours, including $\sim$2 hours overhead for flux, bandpass and complex gain calibration. The resulting uv-coverage for one pointing after 4 $\times$ 2\,min observing time is shown in Fig. \ref{fig_uv_coverage}. The observations were done in two blocks each with 5 hours observing time in April 2012. We chose the quasar 3C286 as a flux and bandpass calibrator and the quasar J1822-0938 as a complex gain calibrator, which was observed every $\sim$13\,min. Observing at L-band and using the new WIDAR correlator, we were able to simultaneously observe the \ion{H}{i} 21cm line, 4 OH lines (1612, 1665, 1667 and 1720 MHz) and 12 H$\alpha$ RRL. For the pilot field, we had the spectral capability to observe 12 H$\alpha$ RRL, however, for the full THOR survey we are able to observe 19 H$\alpha$ RRL. The continuum, consisting of eight spectral windows between 1 and 2\,GHz, was observed in full polarization. For the \ion{H}{i} 21cm line we used a bandwidth of 2\,MHz with a channel width of 1.953\,kHz. This results in a velocity range of $\pm 200$\,km\,s$^{-1}$ and a channel spacing of $\Delta \rm{v} \approx 0.41$\,km\,s$^{-1}$.

\subsection{Calibration}
To edit and calibrate the data, we use CASA (version 4.1.0) with a modified VLA pipeline\footnote{
https://science.nrao.edu/facilities/vla/data-processing/pipeline } (version 1.2.0). The pipeline does automatic flagging for e.g. zeros or shadowing of antennas. Additional flagging for Radio Frequency Interference (RFI) and bad antennas was done manually by hand. The pipeline also applies the bandpass, flux and complex gain calibrator to the data. We do not use Hanning smoothing and do not recalculate the data weights ('statwt') at the end of the pipeline. The implemented modifications help to improve the quality check, and we do some flagging and editing by hand subsequently. A full description of our quality check method will be presented in our forthcoming overview paper (Beuther et al., in prep).

\subsection{Imaging and deconvolution}
The \ion{H}{i} emission and absorption covers a large range of angular scales, which is challenging for the data reduction, as the VLA C-Array filters out most of the large scale structure. Therefore we combined our THOR data with the VGPS data \citep[VLA D-Array and Green Bank Telescope data,][]{Stil2006}, to overcome missing flux problems and to reconstruct the large scale structure. We tested three different methods to combine the THOR and VGPS data. First, we tried to combine the visibility of the VGPS and THOR data and clean them together. Second, we tried the 'feather' command in CASA and third, we used the images of the VGPS survey as a starting model (parameter 'modelimage') for the clean process in CASA. After testing these three methods, we choose the third method, as this method provides the best results considering noise, side lobes and recovery of large scale structure. We compared point source peak fluxes of the combined images to the VGPS data to check for consistency. The difference is on the 10\% level. The clean process was stopped at the 5$\sigma$ noise level.\\
We smoothed over three channels to reduce the noise, resulting in a velocity resolution of $1.24$\,km\,s$^{-1}$. This was the best compromise between computational time, noise and still a sufficient spectral resolution to distinguish absorption/emission features. The synthesized beam was set to 20\arcsec, which is slightly larger than the best resolution achievable ($\sim$16\arcsec). The weighting parameter was set to $robust = 0.5$, which is a combination of natural and uniform weighting. These methods result in an rms of $\sim$14\,K (9mJy beam$^{-1}$) for emission free channels and up to a factor of 2 or 3 more in channels with strong emission, due to systematic errors such as side lobes. The dynamical range of our data set is $\sim$100-200, depending on the region. In the following, the \ion{H}{i} absorption and small scale structure is based on the THOR data, whereas the large scale emission is based on the VGPS data. \\
The 21cm continuum data in this paper are taken from the \ion{H}{i} data cube for high and low velocities (-80 to -50 and 135 to 155\,km\,s$^{-1}$) that are not affected by \ion{H}{i} emission or absorption. Therefore the data reduction and imaging for the 21cm continuum data is the same as for the \ion{H}{i} data and we can avoid systematic errors due to different data reduction methods.

\subsection{H$_{2}$ column density}
\label{observations_H2}
As we will compare the \ion{H}{i} and H$_{\rm{2}}$ column density in Sect. \ref{spatial_distribution_of_hydrogen}, we need an estimate of the H$_{\rm{2}}$ content. We use dust observations from the Herschel Space Observatory to assess the H$_{\rm{2}}$ column density. These observations are based on the HiGal survey \citep{Molinari2010}. The H$_{\rm{2}}$ column density map is taken from Fig. 9 of \citet{Nguyen2013}, which was derived by SED fitting methods described by \citet{Hill2011}. As the dust observations have no velocity resolution, we see all the dust and thus gas along the line of sight. \citet{Carlhoff2013} showed that the Herschel dust data are similar within the uncertainties to the CO data at the velocity range of W43 ($v_{\rm{LSR}} = 60-120$\,km\,s$^{-1}$). Hence, the Herschel dust observations are dominated by the emission from W43 and the contributions of other regions along the line of sight can be neglected. We refrain from using CO data to estimate the molecular content of W43, as CO does not trace all molecular hydrogen \citep[e.g.][]{Smith2014} and CO becomes optically thick for the dense interior of W43 \citep{Carlhoff2013}.

\section{\ion{H}{i} radiative transfer}
\begin{figure}
  \resizebox{\hsize}{!}{\includegraphics{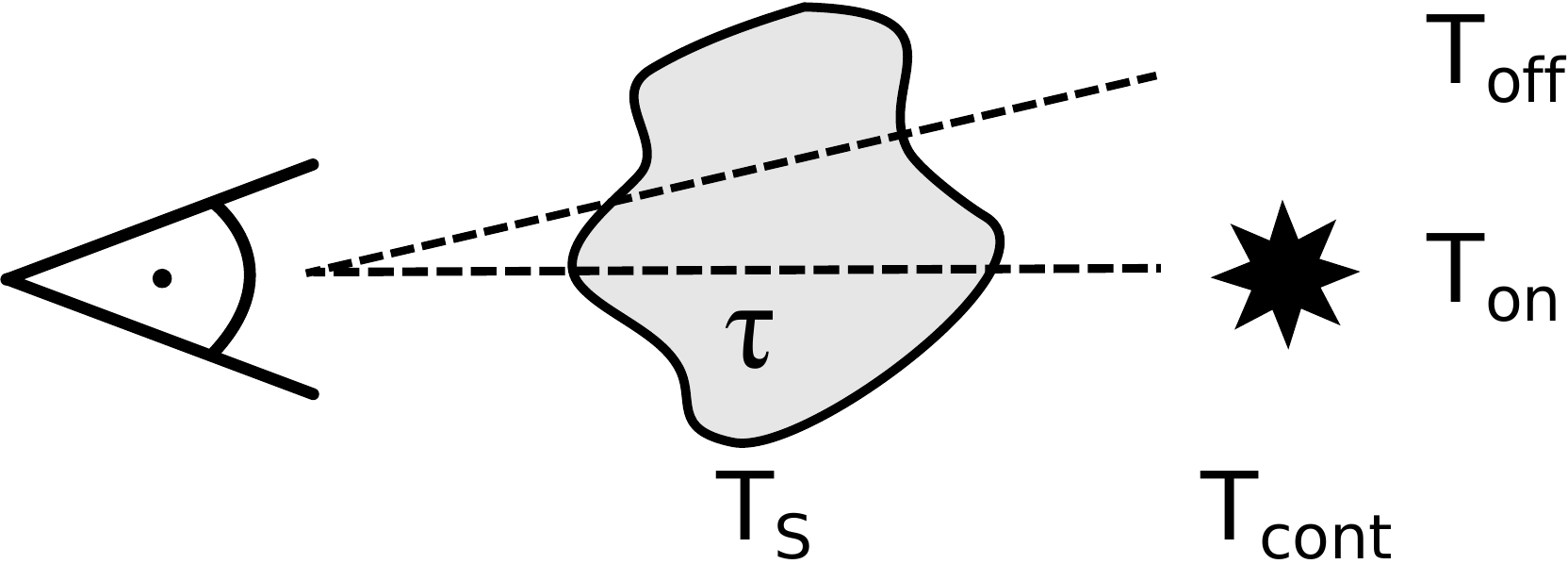}}
  \caption{Sketch of the arrangement of the \ion{H}{i} cloud with spin temperature $T_{\rm{S}}$ and optical depth $\tau$ and the continuum source in the background. The on and off positions are marked.}
  \label{on-off-position-sketch}
\end{figure}
In this section, we explain the methodology we used to determine the spin temperature, optical depth and column density of the neutral hydrogen. Even though the \ion{H}{i} 21cm line is a well known probe of these quantities, the arrangement of different \ion{H}{i} components with different temperatures along the line of sight can complicate these measurements. First we will explain the classical methods to determine the column density via optically thin emission and \ion{H}{i} continuum absorption (HICA). Subsequently, we will outline the limitations of these methods and describe our approach to correct for the optical depth and the weak continuum emission. Although the basics of this approach are discussed in the literature \citep[e.g.,][]{Wilson2010}, we modified the classical approach to account for the continuum emission and optical depth. Therefore, we outline the calculations here in more detail.\\
All following values are frequency dependent. To keep the equations and description simple, we drop the frequency dependencies in the equations. We emphasise however, that the following emission/absorption mechanisms only work for identical frequencies and therefore identical velocities.\\
\subsection{Column density}
The classical approach to determine the \ion{H}{i} column density is given by \citet{Wilson2010} as:
\begin{equation}
\frac{N_{\rm{H}}}{\rm{cm}^{-2}} = 1.8224 \times 10^{18} \: \frac{T_{\rm{S}}}{\rm{K}} \int_{-\infty}^{\infty} \tau(v) \: d\left( \frac{v}{\rm{km\,s}^{-1}}\right) ,
\label{HI_column_density}
\end{equation}
where $N_{\rm{H}}$, $T_{\rm{S}}$ and $\tau$ are the column density, the spin temperature and the velocity dependent opacity, respectively. The spin temperature $T_{\rm{S}}$ describes the relative population of the spin states of the hydrogen atom \citep{Wilson2010}. As $T_{\rm{S}}$ is the equivalent of the excitation temperature for molecules, it is only equal to the kinetic temperature in Local Thermodynamic Equilibrium (i.e. when there are enough collisions to thermalise the gas). We assume in the following that the spin temperature $T_{\rm{S}}$ does not vary significantly within one velocity channel. In most cases, neither $T_{\rm{S}}$ nor the optical depth are known. The simplest assumption is that the \ion{H}{i} emission is optically thin (see Sect. \ref{HI_radiative_trans_optical_thin_HI}).

\subsection{Optically thin \ion{H}{i} emission}
\label{HI_radiative_trans_optical_thin_HI}
Under the optically thin assumption, without background continuum emission, the expression for the brightness temperature $T_{\rm{B}}$ simplifies to the following:
\begin{equation}
T_{\rm{B}} = T_{\rm{S}} \: (1-\rm{exp}(-\tau)) \approx T_{\rm{S}} \: \tau .
\label{optical_thin_HI_emission}
\end{equation}
This simplification provides a linear relation between the column density (equation \ref{HI_column_density}) and the brightness temperature. Hence, we can estimate the column density directly from the measured brightness temperature $T_{\rm{B}}$. This method is used in several studies and is well described in the literature \citep[e.g.,][]{Lee2012, Motte2014, Wilson2010}. Below, we will discuss its limitations and describe a procedure to overcome them.

\subsection{\ion{H}{i} optical depth}
\label{introduction_HICA}
\begin{figure}
  \resizebox{\hsize}{!}{\includegraphics{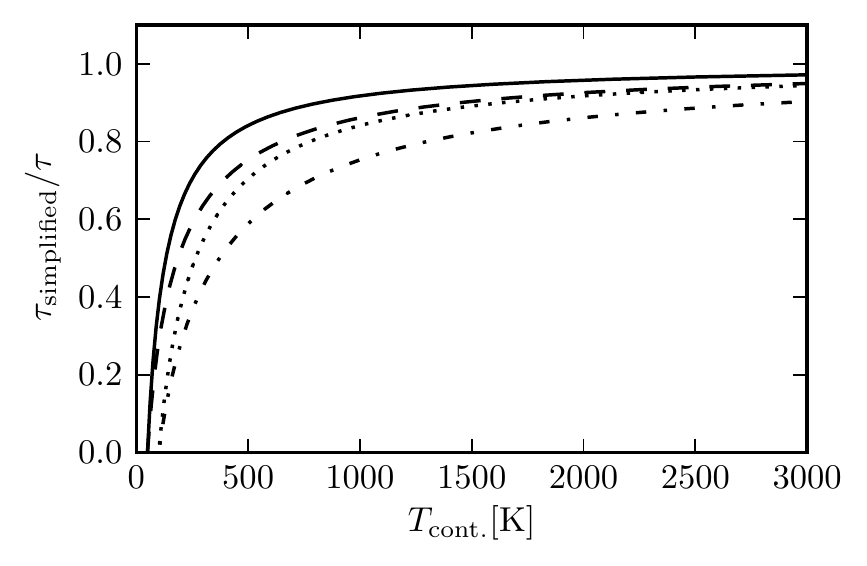}}
  \caption{Optical depth simplification. The simplified optical depth is calculated by neglecting the spin temperature of the cloud. The solid and dashed lines represent a spin temperature of 50\,K with optical depths of 1 and 2, respectively. The dotted and dash-dotted lines represent a spin temperature of 100\,K with optical depths of 1 and 2, respectively.}
  \label{figure_tau_simplified}
\end{figure}

\begin{figure}
  \resizebox{\hsize}{!}{\includegraphics{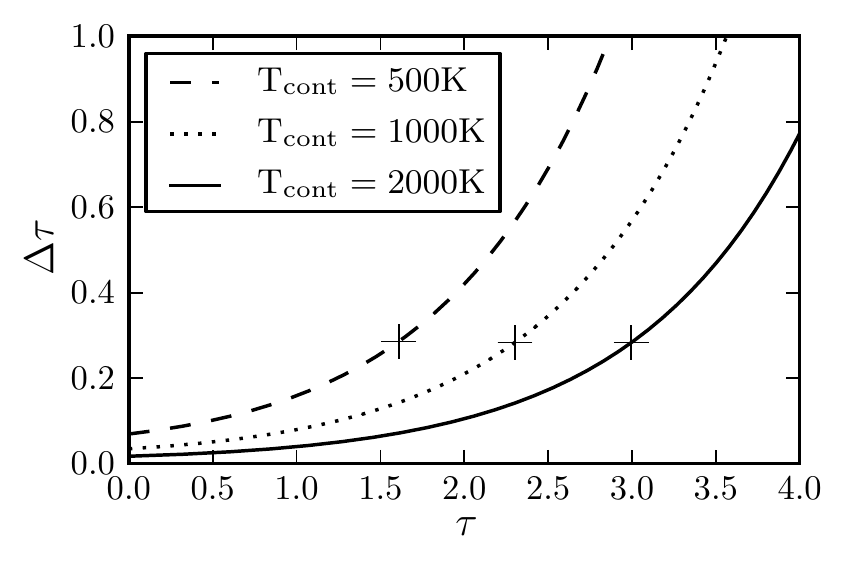}}
  \caption{The absolute uncertainty of the optical depth as a function of the optical depth itself for three different continuum background sources. The dashed, dotted and solid line represent continuum sources with a brightness temperature of $T_{\rm{cont}} = 500$K, 1000K and 2000K, respectively. The crosses on each line show the lower limit of the optical depth that we can observe for the given continuum brightness temperature. For larger optical depths, the absorption spectra saturates (see Sec. \ref{results_optical_depths_determined_using_compact_sources} for more details).}
  \label{figure_optical_depth_uncertainty}
\end{figure}

\begin{figure*}
  \resizebox{\hsize}{!}{
  \includegraphics[width=9cm]{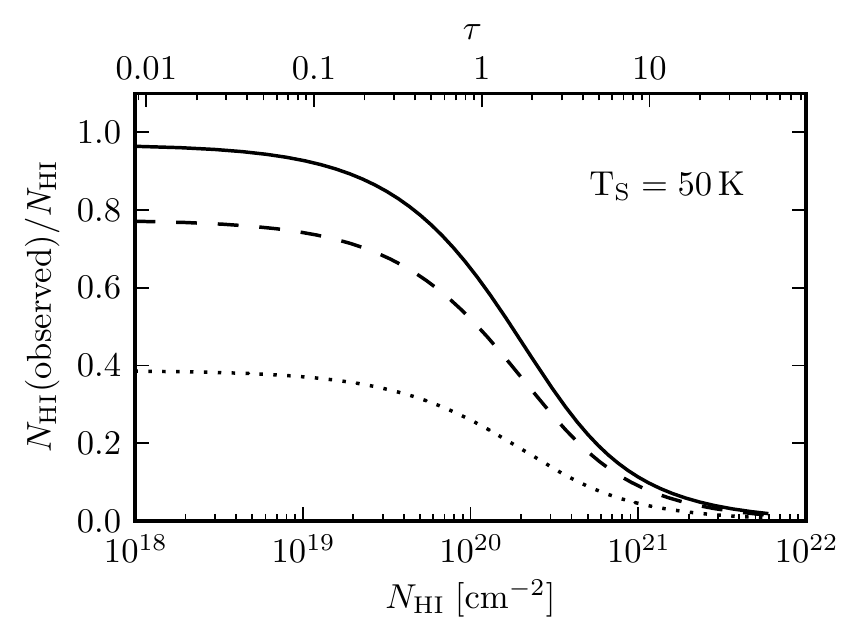}
  \includegraphics[width=9cm]{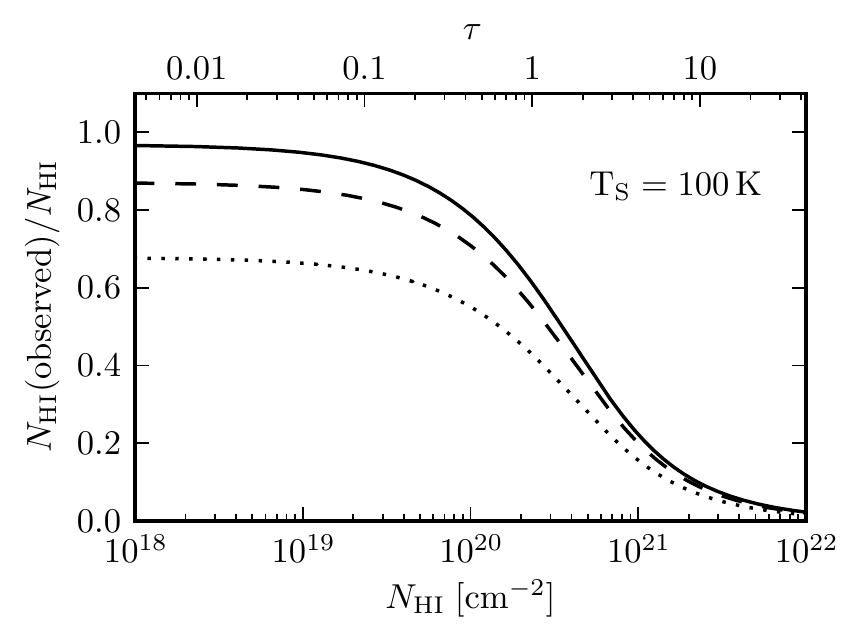}
  }
  \caption{Ratio of observed and expected column density for a model cloud with spin temperatures of $T_{\rm{S}} = 50$\,K and 100\,K for the left and right panel respectively. The solid lines show the ratio for the assumption of optically thin emission. The dashed and dotted lines show the ratio for the assumption of optically thin emission and a diffuse continuum source with brightness 10\,K and 30\,K, respectively.}
  \label{column-density-continuum-corrected}
\end{figure*}

The \ion{H}{i} continuum absorption method is the classical method to determine properties of the cold neutral medium \citep{Strasser2004, Heiles2003a, Heiles2003b, Strasser2007}. This method is based on strong continuum sources ($T_{\rm{B}} > 300$\,K), such as Galactic \ion{H}{ii} regions, Active Galactic Nuclei (AGNs) or extragalactic jets. As the brightness temperature of these continuum sources is larger than the typical spin temperature of the \ion{H}{i} clouds ($T_{\rm{S}} \sim 100$\,K), we observe the \ion{H}{i} cloud in absorption. In addition, the absorption spectrum is dominated by the cold neutral medium, as the \ion{H}{i} absorption coefficient is proportional to $T^{-1}$.\\ 
The classical observing strategy is the `on-off' observation (see Fig. \ref{on-off-position-sketch}). The `on-source' points toward the continuum source and reveals the absorption spectrum ($T_{\rm{on}}$), whereas the `off-source' points slightly offset from the continuum source and reveals the emission spectrum ($T_{\rm{off}}$). During the data reduction, the continuum is not subtracted from the \ion{H}{i} spectrum. Hence, we can measure $T_{\rm{cont}}$ for channels that are not affected by the \ion{H}{i} line. The general assumption is that the on-source and off-source spectra originate from the same cloud with the same properties. Therefore it is important to have these two positions as close together as possible.\\
The measured on-source and off-source brightness temperatures are:
\begin{equation}
\begin{split}
T_{\rm{on}} &= T_{\rm{S}} \: (1-\rm{exp}(-\tau))+T_{\rm{cont}}\:\rm{exp}(-\tau) ,\\
T_{\rm{off}} &= T_{\rm{S}} \: (1-\rm{exp}(-\tau)).
\end{split}
\label{T_on}
\end{equation}
Hence, the optical depth is:
\begin{equation}
\tau = -\rm{ln}\left(\frac{T_{\rm{on}}-T_{\rm{off}}}{T_{\rm{cont}}}\right) .
\label{equation_tau}
\end{equation}
The spin temperature can then be determined from:
\begin{equation}
T_{\rm{S}} = \frac{T_{\rm{off}}}{1-\rm{e}^{-\tau}} .
\label{equation_spin_temperature}
\end{equation}
The advantage of the HICA method is the direct measurement of the optical depth and the spin temperature. The challenge of the HICA method is the need for strong continuum sources. Since most of the strong continuum sources are point sources, it is not possible to map the entire Milky Way, but instead the result is an incomplete grid of measurements. This way, it is not possible to map the intrinsic structure of individual clouds. Furthermore, the spin temperature does not describe the actual temperature of the cloud but is an mean of the cold and warm component, weighted by their column densities \citep{Strasser2004}. Consequently, the derived spin temperature is an upper limit for the cold component.\\
As some Galactic continuum sources, such as W43-Main, are extended, it is difficult to determine a proper off position. Nevertheless, it is still possible to determine the optical depth. As we use the VLA C+D array observations including the continuum without the GBT data, we filter out most smooth large scale structure. The VLA C+D array data show \ion{H}{I} emission of less than 30K. Therefore, we can neglect the emission of the \ion{H}{I} cloud in equation \ref{T_on} and set $T_{\rm{off}} = T_{\rm{S}} \: (1-\rm{exp}(-\tau)) = 0$ and calculate the optical depth without any off position as:
\begin{equation}
\tau_{\rm{simplified}} = -\rm{ln}\left(\frac{T_{\rm{on}}}{T_{\rm{cont}}}\right) .
\label{equation_tau_simplified}
\end{equation}

Even without the effect of filtering out the \ion{H}{I} emission due to the interferometer, this simplification holds for strong continuum sources (T$_{\rm{cont}}>500$K). Figure \ref{figure_tau_simplified} shows the relation between the simplified and actual optical depth as a function of the continuum brightness temperature, not taking into account the filtering of the interferometer. It shows clearly, that even for high spin temperatures ($T_{\rm{S}} = 100$\,K) and high optical depth ($\tau = 2$), we miss less than 10\% of the optical depth for a strong continuum source ($T_{\rm{cont}} \approx 2500$\,K). Therefore, we are able to measure the optical depth for the \ion{H}{ii} region in the center of W43, even though we cannot determine a proper off position. \\
We also consider the effect of optically thick clouds. For these optically thick clouds, the absorption spectra approaches zero. Due to the rms of the spectra, it is also possible that the absorption spectra become negative, which is physically not meaningful. We set the optical depth to a lower limit for all absorption values that are close to zero. If the absorption spectrum $T_{\rm{on}}$ is smaller than 5 times the rms, we set the optical depth to:
\begin{equation}
\tau_{\rm{lower-limit}} = -\rm{ln}\left(\frac{5 \cdot \sigma(T_{\rm{on}})}{T_{\rm{cont}}}\right) .
\label{equation_tau_lower_limit}
\end{equation}
We will show that the optical depth saturates for the main velocity range of W43-Main, which has consequences for our interpretation and conclusions.\\
The uncertainty of the optical depth (equation \ref{equation_tau}) depends on $T_{\rm{on}}$, $T_{\rm{off}}$ and $T_{\rm{cont}}$. To estimate the uncertainty, we assume an uncertainty of 20K for all three quantities. The uncertainty of the optical depth is shown in Fig. \ref{figure_optical_depth_uncertainty} as a function of the optical depth itself for three different continuum brightness values. It shows that the uncertainty increases significantly for increasing optical depths. However, we are not able to measure such high optical depths as the absorption spectra saturates for optically thick clouds and we report lower limits (see equation \ref{equation_tau_lower_limit}). These lower limits are marked with crosses in Fig. \ref{figure_optical_depth_uncertainty}. Up to the lower limit of the optical depth the corresponding uncertainty is $\sim$0.3 for the three different continuum brightness values. For strong continuum sources, such as W43-Main, the uncertainty of the optical depth is $\sim$10\% up to the lower limit of $\sim$3.\\

\subsection{Column density corrections}
\label{introduction_column_density_correction}
In addition to distinct, small continuum sources, such as \ion{H}{ii} regions or extragalactic jets, we also find weak diffuse continuum emission in the Galactic plane. This component has a strength between 10 and 50\,K. Therefore, it is not strong enough to induce absorption features (HICA), but nevertheless it can influence the \ion{H}{i} emission. To overcome this problem, the classical approach is to subtract this weak continuum emission during the data reduction. Hence, we subtract $T_{\rm{cont}}$ in equation \ref{T_on} and we get:
\begin{equation}
\begin{split}
T_{\rm{B}} & = T_{\rm{S}} (1-\rm{exp}(-\tau))+T_{\rm{cont}}\;\rm{exp}(-\tau)-T_{\rm{cont}} \\
& = (T_S-T_{\rm{cont}})\;(1-\rm{exp}(-\tau)).
\end{split}
\label{equation_diffuse_continuum_emission}
\end{equation}
This shows that even if we subtract the continuum from our data, it can still influence the measured brightness temperature. If we neglect the weak continuum emission, our measured \ion{H}{i} emission will be suppressed and thus the calculated column density will be underestimated. Therefore it is important to investigate the influence of the weak diffuse continuum emission.\\
In the following, we will assess the assumption of optically thin emission and the influence of weak diffuse continuum emission on the determination of the measured column density for appropriate model clouds. We assume two model clouds which have a spin temperature of 50\,K and 100\,K, respectively. We use equation \ref{HI_column_density} to calculate the expected column density $N_{\rm{\ion{H}{i}}}$ for different optical depths in one velocity channel ($dv = 1.24$\,km\,s$^{-1}$). Furthermore, we assume the cloud to be optically thin and use equation \ref{optical_thin_HI_emission} to calculate the brightness temperature of the \ion{H}{i} emission. Using this result, we can calculate the column density of the cloud, but this time with the simplification of optically thin emission. Hence, we call it the observed column density $N_{\rm{\ion{H}{i}}}(observed)$. The solid lines in Fig. \ref{column-density-continuum-corrected} show the ratio of the expected and observed column density for a range of optical depths. Obviously, for small optical depths ($\tau < 0.1$) our assumption of optically thin emission is sufficient and we observe more than 90\% of the expected column density. But for larger optical depths ($\tau > 0.1$), we miss a significant fraction of the column density (>40\%).\\
In addition, we add weak continuum emission in the background, which changes the brightness temperature according to equation \ref{equation_diffuse_continuum_emission} and therefore suppresses the \ion{H}{i} emission. Nevertheless, if we still assume optically thin emission and do not consider the weak continuum emission, we can calculate the observed column density. The dashed and dotted lines in Fig. \ref{column-density-continuum-corrected} show this case for continuum emission of 10\,K and 30\,K, respectively. Even for small optical depths ($\tau < 0.1$) we miss a significant fraction of the column density, which depends on the ratio of the spin temperature and the continuum emission brightness temperature. In the worst case ($T_S = 50$\,K and $T_{\rm{cont}}=30$\,K), we observe only 40\% of the expected column density, even for small optical depths.\\
To summarize, we measure the brightness temperature of the \ion{H}{i} emission T$_{\rm{B}}$, as well as the brightness temperature for the continuum T$_{\rm{cont}}$ and combine these information with the optical depth $\tau$, which we measure toward strong continuum sources. This allows us to calculate the corrected column density by combining equation \ref{HI_column_density} and \ref{equation_diffuse_continuum_emission}:
\begin{equation}
N_{\rm{H}} = 1.8224 \times 10^{18} \: \left(\frac{T_{\rm{B}}}{1-e^{-\tau}} + T_{\rm{cont}} \right) \int_{-\infty}^{\infty} \tau(v) \: dv ,
\label{equation_HI_column_density_corrected}
\end{equation}
where the column density N$_{\rm{H}}$ is given in units $\rm{cm^{-2}}$, the brightness temperature T$_{\rm{B}}$ and the continuum brightness temperature $T_{\rm{cont}} $ are measured in K and the velocity dv is given in $\rm{km\,s}^{-1}$.
\begin{figure}
  \resizebox{\hsize}{!}{\includegraphics{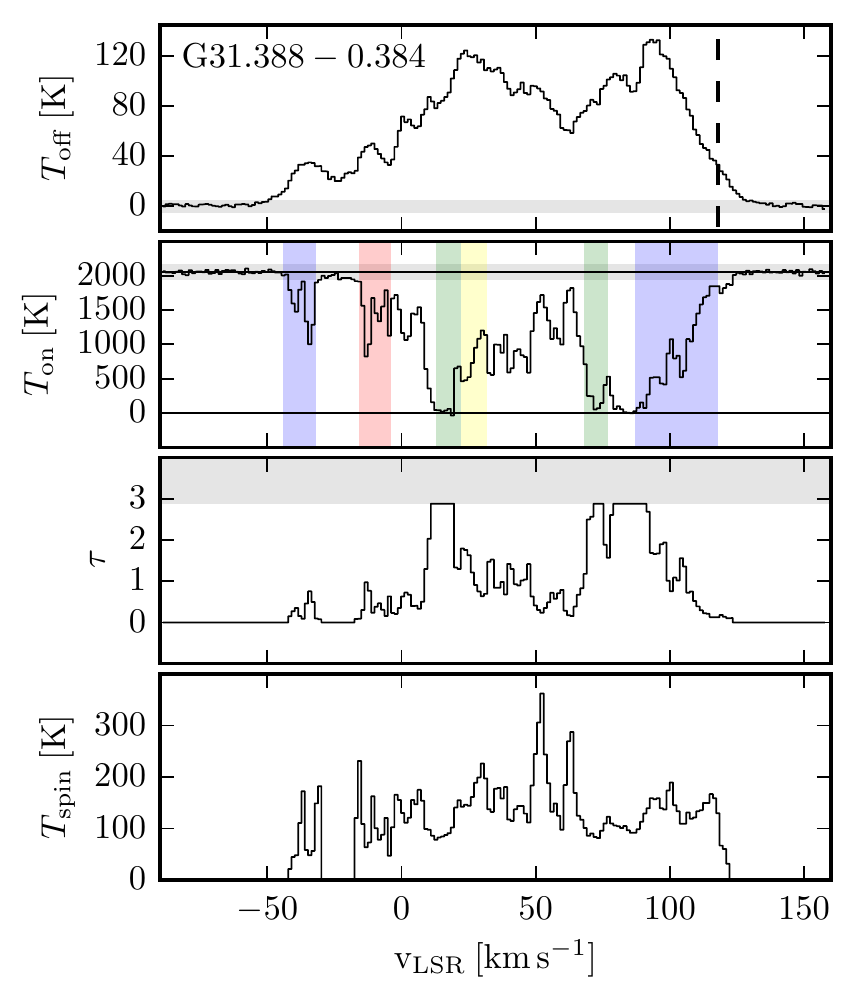}}
  \caption{\ion{H}{i} emission and absorption spectrum of the extragalactic continuum point source G31.388$-$0.384. The emission spectrum is shown in the first panel and is measured in an annulus around the point source with inner and outer radii of 60\arcsec and 120\arcsec, respectively (corresponds to 3 and 6 times the restoring beam). The second panel presents the absorption spectrum toward the point source. The color shaded areas represent the approximate velocities of the Milky Way spiral arms \citep{Vallee2008} in blue, red, green and yellow the Scutum-Centaurus, Cygnus, Sagittarius and Perseus arm, respectively, and the dashed black line in the first panel indicates the tangential velocity. In the first two panels, the grey shaded area indicates the $5\sigma$ noise level. The third panel shows the optical depth computed using equation \ref{equation_tau} and the grey shaded area indicates the saturated optical depth limit, computed using equation \ref{equation_tau_lower_limit}. In the fourth panel the spin temperature is presented, which is computed using equation \ref{equation_spin_temperature_corrected}.}
  \label{figure_continuum_extra_gal}
\end{figure}
\begin{figure}
  \resizebox{\hsize}{!}{\includegraphics{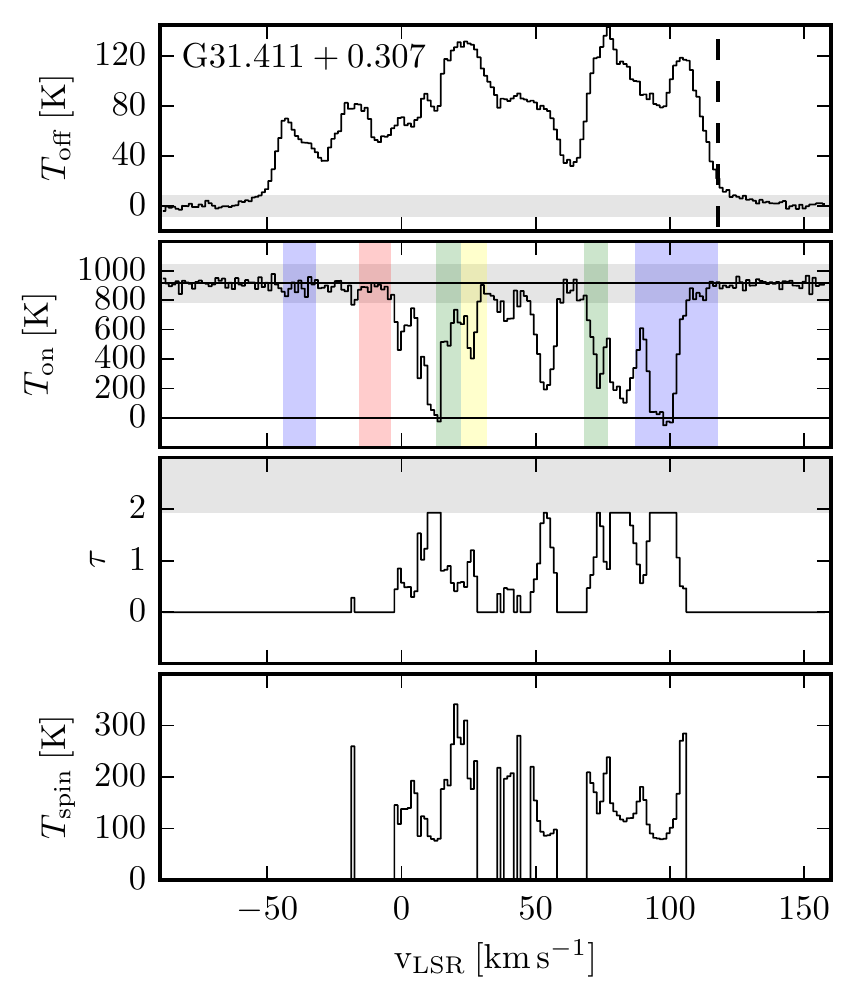}}
  \caption{Same layout as Fig. \ref{figure_continuum_extra_gal}, but for the Galactic continuum source G31.411+0.307}
  \label{figure_continuum_gal}
\end{figure}

\begin{table*}
\caption{Continuum point sources that were used for \ion{H}{i} absorption studies.}             
\label{table_point-sources}      
\centering    
\begin{tabular}{l c c c c c  c c c c} 
\hline\hline
Name & RA.(2000) & Dec.(J2000) & MAJ x MIN & $T_{\rm{cont}}$ & $T_{\rm{cont,dif}}$ & far Scutum  & spectral & location   \\
 & (h : m : s) & $(^{\circ}:\arcmin:\arcsec)$ & ($\arcsec$ x $\arcsec$) & K & K & arm & index\\
\hline  
G31.388$-$0.384 & 18:49:59.1 & -1.32.55.9 & 21.3 x 20.3 &  2055$\pm$24 & 20.2 & yes  & -0.82 & ext. gal\\
G31.411+0.307 & 18:47:34.1 & -1.12.45.5 & 24.0 x 23.0 &  914$\pm$26 & 19.4 & no & 0.63 & gal.\\
G30.234$-$0.138 & 18:47:00.4 & -2.27.52.9 & 21.7 x 20.7 &  853$\pm$23 & 42.1 & no & 0.00 & gal.\\
G30.534+0.021 & 18:46:59.3 & -2.07.27.9 & 28.0 x 21.5\tablefootmark{a} &  653$\pm$21 & 35.9 & no  & 0.21  & gal.\\
G31.242$-$0.110 & 18:48:44.7 & -1.33.13.7 & 25.1 x 21.2 &  573$\pm$22 & 30.1 & no & 0.58 & gal.\\
G29.090+0.512 & 18:42:35.8 & -3.11.04.7 & 20.0 x 10.0 &  571$\pm$44 & 20.5 & yes & -1.03 & ext. gal\\
G30.744+1.008 & 18:43:51.4 & -1.29.14.8 & 23.1 x 20.1 &  472$\pm$36 & 16.9 & yes & -0.90 & ext. gal\\
G30.699$-$0.630 & 18:49:36.3 & -2.16.27.2 & 22.3 x 19.6 &  403$\pm$18 & 23.4 & yes & -1.11 & ext. gal\\
\hline   
\end{tabular} 
\tablefoot{Continuum point sources in the observed field with brightness temperatures $T_{\rm{cont}}>400$\,K. The names and spectral indexes are based on work presented in Johnston et al. (in prep) and are preliminary results. Coordinates, major and minor axes of the 2D fit, maximum brightness temperature $T_{\rm{cont}}$ of the continuum point source and the diffuse weak continuum temperature around the point source are given. The column `far Scutum arm' indicates if absorption features at $v_{\rm{LSR}} \approx -40$\,km\,s$^{-1}$ are present in the \ion{H}{i} spectra. The spectral index and the absorption of the far Scutum arm are used to determine the location of the point sources, which is given in the last column.\\
\tablefoottext{a}{This source consists of two blended sources, which explains the large eccentricity.}\\
}
\end{table*}

\subsection{Continuum correction for strong point sources}
\label{section_continuum_correction_for_ps}
For the \ion{H}{i} continuum absorption (HICA) toward strong continuum sources, we also have to consider weak diffuse continuum emission, which contributes to the on and the off position. Therefore we have to modify the equation for the on and off position (equation \ref{T_on}) by adding another term for the weak diffuse continuum emission, which we call $T_{\rm{cont, dif}}$. This does not change our result for the optical depth (equation \ref{equation_tau}), as the optical depth depends solely on the difference between $T_{\rm{on}}$ and $T_{\rm{off}}$. However, for the calculated spin temperature, we have to modify equation \ref{equation_spin_temperature} to:
\begin{equation}
T_{\rm{S}} = \frac{T_{\rm{off}}-T_{\rm{cont, dif}}\;e^{-\tau}}{1-e^{-\tau}} .
\label{equation_spin_temperature_corrected}
\end{equation}
The effect of the diffuse continuum correction will be discussed in Sect. \ref{discussion_spin_temp}.

\section{Results}

\subsection{Optical depths determined using compact continuum sources}
\label{results_optical_depths_determined_using_compact_sources}

As described in Sect. \ref{introduction_HICA}, we can use strong continuum sources in the background as light-houses, that shine through foreground \ion{H}{i} clouds and create absorption spectra. Below, we characterize the continuum sources.\\ 
To determine the optical depth of the \ion{H}{i} accurately, we need continuum sources in the background which are brighter than the typical \ion{H}{i} spin temperature (see Sect. \ref{introduction_HICA}). We extract all continuum sources in our field that have a brightness temperature $T_{\rm{cont}}>400$\,K and a point-like structure, which yields in 8 point sources. The analysis of extended sources will follow in Sect. \ref{extended_continuum_sources}. We use a two dimensional Gaussian to fit the position and size of the continuum source. To measure the continuum brightness temperature, we average over high and low velocity channels in our \ion{H}{i} data cubes (-80 to -50 and 135 to 155\,km\,s$^{-1}$), that are not affected by the \ion{H}{i} line. The results are shown in Table \ref{table_point-sources}.\\
To determine the off spectrum ($T_{\rm{off}}$), we average the emission spectrum around the continuum source within an annulus with inner and outer radius of 60\arcsec and 120\arcsec, respectively. The upper two panels of Fig. \ref{figure_continuum_extra_gal} show a typical emission and absorption spectrum. We use equation \ref{equation_tau} to calculate the optical depth for each channel. To avoid unrealistic optical depths, we only calculate the optical depth for those channels where the emission/absorption is 5 times larger than the corresponding noise. The grey shaded areas in the upper three panels in Fig. \ref{figure_continuum_extra_gal} and \ref{figure_continuum_gal} show the 5 sigma level. The optical depth is set to zero for channels, where the emission/absorption is below 5 times the corresponding noise.\\ 
As explained in Sec. \ref{introduction_HICA}, we have to consider the effect of optically thick clouds for which we can determine only lower limits for the optical depth. The grey shaded areas in the third panels of Fig. \ref{figure_continuum_extra_gal} and \ref{figure_continuum_gal} show these lower limits of the optical depth. \\

\subsection{\ion{H}{i} spin temperature toward compact continuum sources}
\label{results_HI_spin_temp}
\begin{figure}
  \resizebox{\hsize}{!}{\includegraphics{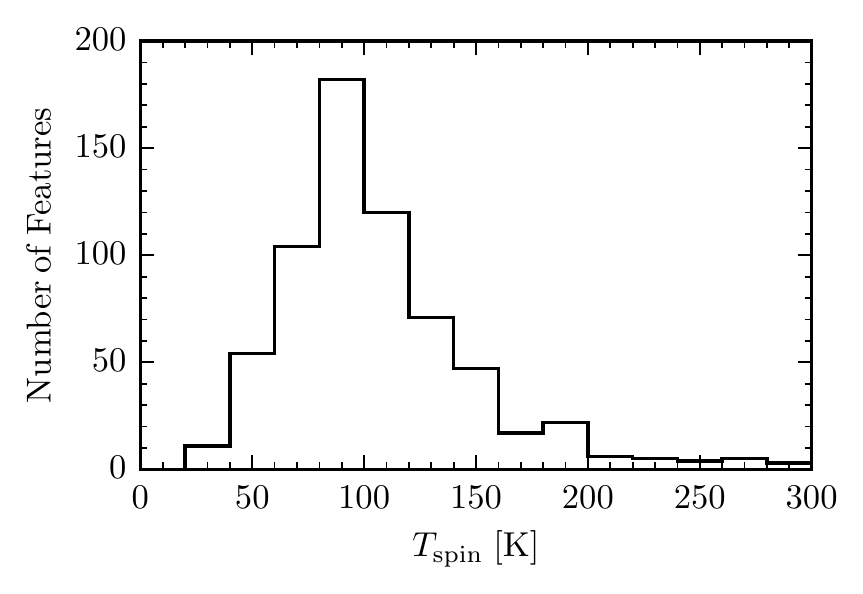}}
  \caption{Distribution of the spin temperature for all absorption features (655) toward point-like continuum sources. The median of the spin temperature is 97.5\,K.}
  \label{figure_histogram_spin_temp}
\end{figure}

For each channel that allows for an optical depth measurement, we compute the spin temperature using equation \ref{equation_spin_temperature_corrected}. The spin temperature is shown in the fourth panel of Fig. \ref{figure_continuum_extra_gal} and \ref{figure_continuum_gal}. This method reveals absorption features in 655 channels for all eight continuum sources. The median of the spin temperature is 97.5\,K and the distribution of all absorption features is shown in Fig. \ref{figure_histogram_spin_temp}. 

For large optical depths, the spin temperature approaches the brightness temperature of the off-position. However, for large optical depths ($\tau \gtrsim 2$) we can only report lower limits for $\tau$. Hence, we overestimate the spin temperature in the optically thick regime. Nevertheless this overestimation is small, as for optical depths of $\tau = 2$ the measured spin temperature is $T_{\rm{spin}} \approx 1.16 \cdot T_{\rm{off}}$ (see equation \ref{equation_spin_temperature}). Therefore we overestimate $T_{\rm{spin}}$ by at most $\sim 15\%$. We will discuss these results in detail in Sect. \ref{discussion_spin_temp}.

\subsection{Location of continuum sources}
\label{results_location_of_continuum_sources}
To characterize and understand the \ion{H}{i} absorption spectrum toward the continuum point sources, it is important to know the location of the continuum source. We can distinguish between extragalactic and Galactic point sources. The Galactic sources are mostly \ion{H}{ii} regions, whereas the extragalactic point sources are radio lobes from extragalactic jets or AGN. To distinguish between them, we have two criteria: the spectral index and \ion{H}{i} absorption of the far Scutum-Centaurus arm. As we have to consider e.g. primary beam effects or different spatial filtering of the interferometer for different frequencies, it is very difficult to determine accurate spectral indexes \citep[e.g.][]{Rau2014, Bhatnagar2011, Bhatnagar2013}. For this analysis, the spectral indices were calculated using flux measurements in the two most-separated spectral windows at 1.05 and 1.95\,GHz and are based on work presented in Johnston et al. (in prep.).\\
Four sources in our sample have negative spectral indices (Table \ref{table_point-sources}), which is typical of synchrotron radiation from extragalactic jets or AGNs. The remaining four sources show flat or positive spectral indices, which indicates free-free emission from potentially optically thick \ion{H}{ii} regions. These four free-free emission sources also show no \ion{H}{i} absorption feature for the far Scutum-Centaurus arm. Therefore, these sources reside in the Milky Way. Furthermore, we can study the \ion{H}{i} absorption spectra and molecular emission spectra of the Galactic point sources to estimate their distance and overcome the near-far distance ambiguity. This is done by \citet{Anderson2014} and they find that e.g. the source G31.411+0.307 has a near distance of 6.6\,kpc. For this source we also see a sharp cut off in the absorption spectra at $v_{\rm{lsr}} \approx 100$\,km\,s$^{-1}$ (see Fig.\ref{figure_continuum_gal}), which corresponds to the molecular line velocity reported in \citet{Anderson2014}.\\
Figure \ref{figure_continuum_extra_gal} shows an example of an extragalactic continuum source. The characteristic absorption of the far Scutum-Centaurus arm at $v_{\rm{lsr}} \approx -40$\,km\,s$^{-1}$ is clearly observable. Furthermore, the absorption and emission spectra show a similar cut off for high velocities, when approaching the tangential point of the near Scutum-Centaurus arm. On the other hand, the absorption spectrum of Fig. \ref{figure_continuum_gal} neither shows the absorption of the far Scutum-Centaurus arm, nor approaches zero at the velocity of the tangential point. This is typical for Galactic continuum sources.\\

\subsection{Extended continuum sources - W43-Main}
\label{extended_continuum_sources}

\begin{figure}
  \resizebox{\hsize}{!}{\includegraphics{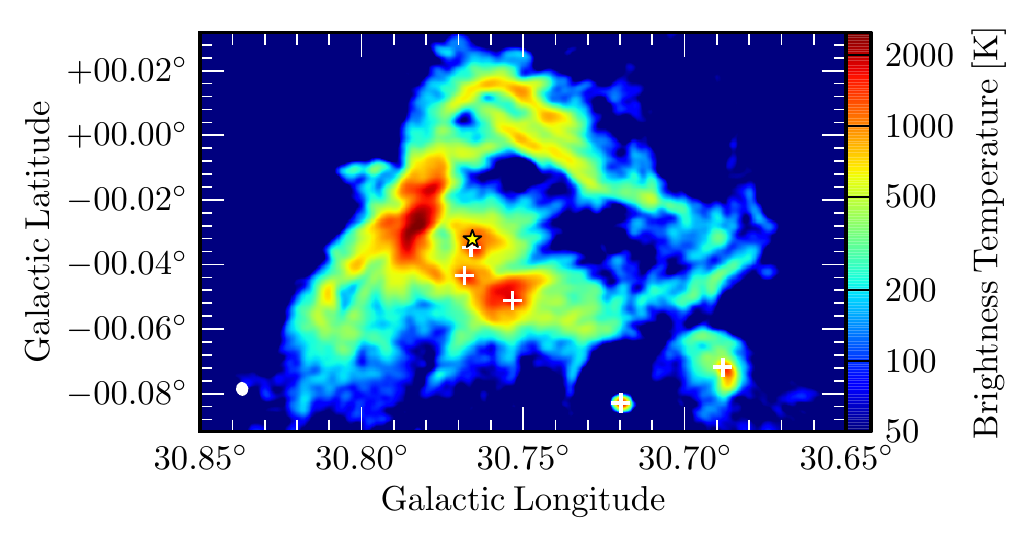}}
  \caption{1.4\,GHz continuum emission of W43-Main. The yellow star indicates the position of the OB-cluster \citep{Lester1985} and the white crosses marks the position of UC\ion{H}{II} regions observed by the CORNISH survey \citep{Hoare2012,Purcell2013}}
  \label{figure_W43-Main-continuum}
\end{figure}

Besides point-like continuum sources, our observed field contains three strong ($T>800$\,K) extended continuum sources. The most prominent source is the well known \ion{H}{ii} region around W43-Main \citep{Lester1985}, which is shown in Fig. \ref{figure_W43-Main-continuum}. This region has an angular extent of $\sim 300$\arcsec, which corresponds to $\sim 8$\,pc at a distance of 5.5\,kpc. As our resolution is 20\arcsec, we are able to resolve the internal structure well. It is known that the nebula is ionized by an OB-cluster and the observed continuum signal is the result of free-free emission. Figure \ref{figure_W43-Main-continuum} shows the continuum emission of W43-Main and the yellow star marks the position of the OB-cluster \citep[J2000, 18h47m36s, -1\degr 56\arcmin 33\arcsec,][]{Lester1985}. Furthermore, several UC\ion{H}{II} regions can be found in the CORNISH survey, which are marked with white crosses \citep{Hoare2012,Purcell2013}. The maximum brightness temperature for this region is $T_{\rm{cont}} \sim 2200$\,K and therefore we are able to calculate the optical depth, even though we cannot determine a proper off position (see Sect. \ref{introduction_HICA} for details). The \ion{H}{i} absorption spectra shows a cut-off at $v_{\rm{lsr}} \sim 100$\,km\,s$^{-1}$ (see Fig. \ref{figure_W43_optical_depth}), which marks the velocity of the continuum source. We measure radio recombination line emission at the same velocity \citep[Johnston et al. in prep, ][]{Anderson2011}. Therefore, the continuum source W43-Main is situated at $v_{\rm{lsr}} \approx 100$\,km\,s$^{-1}$, which is close to the tangent point velocity of the Scutum-Centaurus arm \citep{Nguyen2011}.\\
Furthermore, the observed field contains two other extended continuum sources: W43-South and the supernova remnant SNR G029.7-00.2 with continuum brightness temperatures of at most $T_{\rm{cont}} \approx 1840$\,K and $T_{\rm{cont}} \approx 850$\,K, respectively.\\

\subsection{Optical depth of W43-Main}

\begin{figure}
  \resizebox{\hsize}{!}{\includegraphics{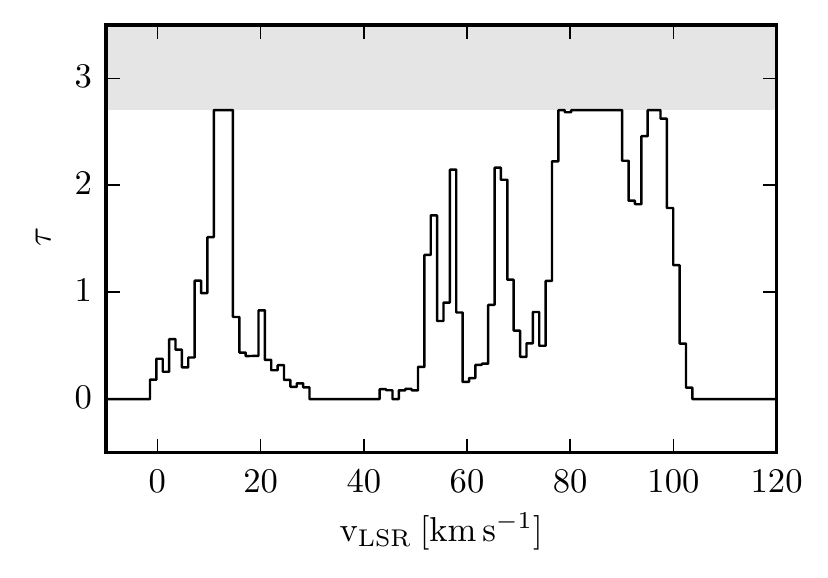}}
  \caption{Optical depth of W43-Main, calculated using equation \ref{equation_tau_simplified}. The grey shaded area indicates the maximum measurable optical depth of $\tau_{\rm{lower-limit}} = 2.7$ calculated using equation \ref{equation_tau_lower_limit}.}
  \label{figure_W43_optical_depth}
\end{figure}

In Sect. \ref{introduction_HICA}, we described that we can determine the optical depth for strong continuum sources by using the absorption spectrum. As the brightness of the continuum source W43-Main is $T_{\rm{cont}} \sim 2200$\,K, the uncertainty for the optical depth measurement is $\sim 10\%$ (see Fig. \ref{figure_tau_simplified}). Figure \ref{figure_W43_optical_depth} shows the optical depth of W43-Main and the grey shaded area indicates the lower limit of our measurement with $\tau_{\rm{lower-limit}} = 2.7$. The optical depth peak at $v_{\rm{lsr}} \approx 10$\,km\,s$^{-1}$ can be allocated to the near Sagittarius arm and therefore is not connected to the actual star forming region W43. In contrast to that, it is difficult to allocate the distinct absorption features between 50 and 80\,km\,s$^{-1}$ and it is not clear whether they are spatially connected to W43. The prominent absorption feature of W43 is situated between 80 and 100\,km\,s$^{-1}$. In this region, our measurement is saturated and therefore we can only report lower limits for the optical depth.\\
As we resolve the strong continuum source W43-Main spatially, we can determine the optical depth along different lines of sight and thus investigate the spatial distribution of absorption features as done in detail by \citet{Liszt1993}. We refrain from such a study as we are mostly interested in the velocity range of W43 ($v_{\rm{lsr}} = 80-110$\,km\,s$^{-1}$) at which the absorption features are mostly saturated preventing a detailed study of the spatial distribution. We instead measured the optical depth toward the strongest continuum peak to estimate the maximum optical depth possible (see Fig. \ref{figure_W43_optical_depth}), which is nevertheless a lower limit.\\

\subsection{\ion{H}{i} column density with optically thin assumption}
\begin{figure*}
   \resizebox{\hsize}{!}
            {\includegraphics[width=18cm]{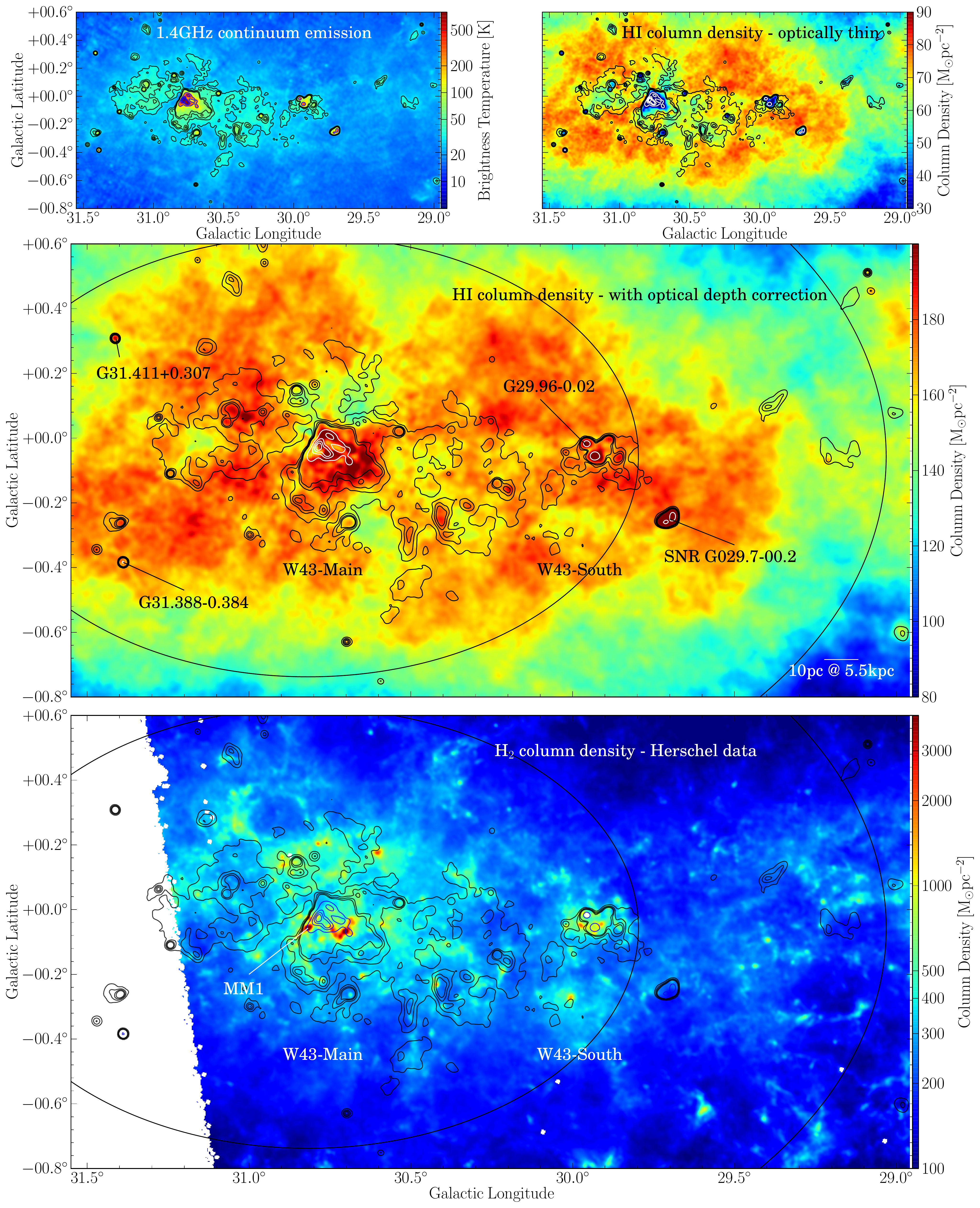}}
      \caption{The top-left panel shows the continuum emission at 21cm in Kelvin. The top-right and middle panels presents the \ion{H}{i} column density for optically thin assumption and optical depth corrections, respectively. The continuum and \ion{H}{i} emission data are based on VGPS, whereas the optical depth correction used in the middle panel is based on the THOR data. The bottom panel shows the H$_{\rm{2}}$ column density from the HiGAL data \citep{Nguyen2013}. In all panels, the black and white/blue contours present the continuum emission at 21cm (black contours show levels of 10, 30 and 70\,K, white/blue contours show levels of 200, 400, 600 and 800\,K). The black ellipses in the middle and lower panel have an equivalent radius of $r=80$ and 140\,pc. Several important objects are marked.}
         \label{figure_HI_column_density_comparison}
\end{figure*}

In this section we assess the \ion{H}{i} column density and the \ion{H}{i} mass of the GMC associated with W43. These two quantities depend strongly on the chosen velocity range. \citet{Nguyen2011} defined for the `main' and `complete' velocity range of W43 values of $v_{\rm{lsr}}(main) = 80-110$\,km\,s$^{-1}$ and $v_{\rm{lsr}}(complete) = 60-120$\,km\,s$^{-1}$. In the following, we will use the `complete' velocity range, but we stress that a different velocity range can change the column density and mass significantly.\\
As described in Sect. \ref{HI_radiative_trans_optical_thin_HI}, we can determine the column density by assuming optically thin \ion{H}{i} emission. For this method, the column density is proportional to the observed \ion{H}{i} brightness temperature. The column density map assuming optically thin emission is shown in the top-right panel of Fig. \ref{figure_HI_column_density_comparison}. To estimate the mass, we integrate this column density over the main region of W43 ($l = 29.0-31.5$\degr, $|b| \leq 1$\degr). We mask all regions, where the emission spectrum has negative values, and exclude them. Given a distance of 5.5\,kpc \citep{Zhang2014}, our analysis finds an \ion{H}{i} mass of $M = 2.7 \times 10^6$\,M$_{\odot}$ ($l = 29.0-31.5$\degr, $|b| \leq 1$\degr, $v_{\rm{lsr}} = 60-120$\,km\,s$^{-1}$).\\
Weak diffuse continuum emission can influence the \ion{H}{i} emission spectrum and therefore the \ion{H}{i} column density calculation needs to be modified according to equation \ref{equation_diffuse_continuum_emission} (Sect. \ref{introduction_column_density_correction}). The top-right panel of Fig. \ref{figure_HI_column_density_comparison} shows this effect clearly. The color represents the column density determined by the optically thin assumption and the black contours indicate the weak diffuse continuum emission. These two components show a clear anti-correlation. However, this anti-correlation is the result of the expected observational effect in which \ion{H}{i} emission is suppressed by weak continuum emission (see Sec. \ref{introduction_column_density_correction}). To overcome this problem we have to consider the optical depth, which we discuss in the next section.\\

\subsection{\ion{H}{i} column density with optical depth correction}
\label{results_column_den_with_correction}
With the measurement of the optical depth and the weak continuum emission, we can correct the \ion{H}{i} emission as explained in Sect. \ref{introduction_column_density_correction}. This allows a more accurate determination of the column density, which is shown in the middle panel of Fig. \ref{figure_HI_column_density_comparison}. We chose the `complete' velocity range of $v_{\rm{lsr}} = 60-120$\,km\,s$^{-1}$. As we correct for the optical depth, we are able to observe a larger \ion{H}{i} column density. We also correct for the weak continuum emission around W43-Main. This correction removes the anti-correlation between the continuum emission and the \ion{H}{i} column density. Hence, we determine the \ion{H}{i} mass to be $M = 6.1 \times 10^6$\,M$_{\odot}$ ($l = 29.0-31.5$\degr, $|b| \leq 1$\degr, $v_{\rm{lsr}} = 60-104$\,km\,s$^{-1}$).\\
The optical depth spectrum of W43-Main (Fig. \ref{figure_W43_optical_depth}) shows, that the absorption ends abruptly at 100\,km\,s$^{-1}$, but the \ion{H}{i} emission as well as CO emission \citep{Nguyen2011, Carlhoff2013} reveals features up to $v_{\rm{lsr}} = 110$\,km\,s$^{-1}$. The reason for the abrupt drop in the absorption spectrum is not the absence of \ion{H}{i}, but the location along the line of sight of the continuum source at this velocity (see Sect. \ref{results_location_of_continuum_sources} for further details). As we do not see \ion{H}{i} in absorption for velocities larger than 100\,km\,s$^{-1}$, we are also not able to determine the corresponding optical depth. Therefore we cannot apply our corrections to the \ion{H}{i} column density for velocities larger than 100\,km\,s$^{-1}$. Hence, the velocity range for the previously mentioned \ion{H}{i} mass is only $v_{\rm{lsr}} = 60-104$\,km\,s$^{-1}$. Nevertheless we can determine the \ion{H}{i} mass for the velocity range from $v_{\rm{lsr}} = 104-120$\,km\,s$^{-1}$ by using the optically thin assumption. This reveals a \ion{H}{i} mass of $M = 0.5 \times 10^6$\,M$_{\odot}$. Hence, the total \ion{H}{i} mass for W43 in the velocity range $v_{\rm{lsr}} = 60-120$\,km\,s$^{-1}$ is $M = 6.6 \times 10^6$\,M$_{\odot}$ with the optical depth correction for the velocity range $v_{\rm{lsr}} = 60-104$\,km\,s$^{-1}$ and optically thin assumption for the velocity range $v_{\rm{lsr}} = 104-120$\,km\,s$^{-1}$. This is 2.4 times larger, than the \ion{H}{i} mass determined with the optically thin assumption.\\
The limitations and uncertainties of our determined \ion{H}{i} column density and \ion{H}{i} mass with the optical depth corrections will be further discussed in Sect. \ref{discussion_column_density}. 

\subsection{\ion{H}{i} self absorption}
\label{results_HISA}
The \ion{H}{i} Self Absorption (HISA) method uses the diffuse broad \ion{H}{i} emission background of the Milky Way as illumination for colder foreground clouds \citep[e.g.][]{Gibson2000,Gibson2005,Gibson2005b,Li2003,McClure-Griffiths2006}. Dark regions on maps and narrow absorption features in spectra reveal these HISA features. The terminology `self absorption' can be misleading, as the \ion{H}{i} emission and absorption can occur in the same cloud or at the same position but does not have to. The advantage of this method is that it is possible to map entire absorption clouds and study their intrinsic structure. On the other hand, the disadvantage is that a sufficient background emission with the same velocity as the absorbing cloud is necessary to detect HISA features. Therefore this method misses a large portion of the cold \ion{H}{i} clouds. The differentiation between actual HISA features and the lack of \ion{H}{i} emission can also be challenging. \citet{Gibson2005b}, however, present an efficient method to detect HISA features automatically. Another disadvantage is that the optical depth and spin temperature can only be measured together and further assumptions are needed to disentangle these two values.\\
As described in the previous section, we correct for the optical depth effects and weak continuum emission. This correction does not account for locally confined HISA features, as we assume a uniform optical depth for the entire W43 region. HISA features could have a higher and spatially varying optical depth that we cannot measure. Furthermore, the weak and diffuse continuum emission around W43 makes the search for HISA features even more complicated. Hence, we refrain from searching and analyzing the possible HISA features around W43. Therefore we are likely missing some cold neutral hydrogen in our analysis. HISA studies will be conducted in other parts of the Milky Way THOR study with less diffuse continuum emission.

\section{Discussion}

\subsection{Phases of the Neutral Atomic Hydrogen}
\label{discussion_CNM_WNM}

\begin{figure}
  \resizebox{\hsize}{!}{\includegraphics{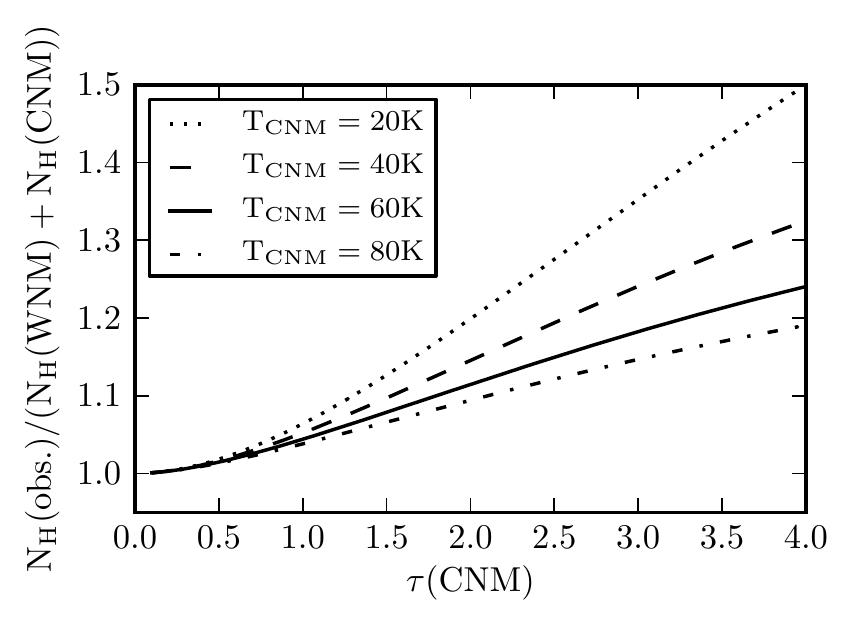}}
  \caption{The ratio of the observed and actual column density as a function of the CNM optical depth $\tau$(CNM) for different CNM spin temperatures T$_{\rm{spin}}$(CNM).}
  \label{figure_column_density_comparison_of_cnm_wnm}
\end{figure}

It is well known that the neutral atomic ISM has several phases that coexist side by side with very different properties \citep[e.g.][]{Clark1965, Wolfire1995, Heiles2003b, Wolfire2003}. The main constituents are the cold neutral medium (CNM) and the warm neutral medium (WNM) with spin temperatures on the order of <100K \citep{Strasser2007} and $\sim10^4$K \citep{Murray2014, Roy2013}, respectively. Furthermore, their density differs by two order of magnitude \citep[CNM: n$_{\rm{H}}\sim50$cm$^{-3}$, WNM: n$_{\rm{H}}\sim0.5$cm$^{-3}$, e.g.,][]{Stahler2004}. Due to the different spin temperature and density of the CNM and WNM, their corresponding optical depths are significantly different with typical values of $\tau_{\rm{WNM}} \sim 10^{-3}-10^{-4}$ \citep{Murray2014} and $\tau_{\rm{CNM}} \gtrsim 0.1$ \citep{Strasser2004}. This is important for our interpretation.\\
Due to the low optical depth of the WNM, we see in absorption merely the CNM. Hence, the optical depth shown in Fig. \ref{figure_W43_optical_depth} is the optical depth of the CNM. As the absorption spectra are strongly dominated by the CNM, we are not able to measure the optical depth of the WNM and we can assume the WNM to be optically thin. In contrast to this, we see a combination of the CNM and WNM for the \ion{H}{i} emission. For the correction of the column density (see Sec. \ref{results_column_den_with_correction}) we use the optical depth information from the absorption study to correct the \ion{H}{i} emission data. As we do not distinguish between the two phases, we might combine two different quantities, namely the optical depth of the CNM with the emission of the CNM and the WNM. This could lead to an overestimation of the column density. In the following, we will assess this effect.\\
If we assume a CNM cloud with varying T$_{\rm{spin}}$(CNM)$\sim$20-80K and varying optical depth surrounded by a WNM with T$_{\rm{spin}}$(WNM)$\sim$7000K \citep{Murray2014} and optical depth $\tau$(WNM)$\sim$5$\times$10$^{-3}$ we can calculate the column density of each component separately (CNM and WNM). Furthermore, we can calculate the brightness temperature that we would observe and apply our correction method described in Sec. \ref{introduction_column_density_correction}. Finally, we can compare the actual column density of the CNM and WNM with the column density we would observe with our correction method and investigate the overestimation of the actual column density. The result is shown in Fig. \ref{figure_column_density_comparison_of_cnm_wnm}, which shows the ratio of the observed and actual column density as a function of the optical depth of the CNM $\tau$(CNM) for different CNM spin temperatures T$_{\rm{spin}}$(CNM). Fig. \ref{figure_column_density_comparison_of_cnm_wnm} shows that we measure the column density accurately for the optically thin case and we overestimate the column density for larger optical depths. However, even for the extreme case (T$_{\rm{spin}}$(CNM)$\sim$20K, $\tau$(CNM)$\sim$3), we overestimate the column density by at most 1.35. This effect is smaller than the underestimation of the column density due to saturated optical depths and therefore we will consider only a single cold component in the following. Similar results in simple simulations for different combination of CNM and WNM fraction and a wide range of N$_{H}$ and $\tau$ were found by \citet{Chengalur2013} and \citet{Roy2013,Roy2013b}.\\

\subsection{\ion{H}{i} spin temperature measurements toward point sources}
\label{discussion_spin_temp}
The spin temperature is the best measure of the kinetic temperature of the \ion{H}{i} gas along the line of sight. Hence, it is a principal determinant of the physical processes that occur in the \ion{H}{i} gas.\\
As described in Sect. \ref{section_continuum_correction_for_ps}, we corrected the \ion{H}{i} spin temperature for diffuse weak continuum emission. If we neglect this correction we would systematically overestimate the \ion{H}{i} spin temperature and our sample would have a median \ion{H}{i} spin temperature of 110\,K. That means the median value would be $\sim 12$\,K higher compared to our corrected value of 97.5\,K (see Fig. \ref{figure_histogram_spin_temp}).\\
Similar studies of the spin temperature with similar methods can be found in the literature. For example \citet{Strasser2004} report a median spin temperature of 120\,K. This is $\sim 20$\,K higher than our value of 97.5\,K, but they do not consider the diffuse weak continuum emission and therefore probably overestimate the spin temperature.\\
\citet{Heiles2003a, Heiles2003b} developed an extensive method to fit Gaussian components to the absorption and emission spectra. Spectral features shown in absorption and emission are assigned to the cold neutral medium (CNM), whereas emission-only features are assigned to the warm neutral medium (WNM). Using this method, they are able to distinguish these two phases and measure their properties such as as the spin temperature individually. However, they report that it is difficult to use their method for sources close to the Galactic plane ($|b|<10\deg$), as multiple components can overlap and the Galactic rotation can broaden their profiles. Figure \ref{figure_continuum_extra_gal} and \ref{figure_continuum_gal} illustrate this problem for our region and for these spectra it is impossible to find unique Gaussian components. As \citet{Heiles2003a, Heiles2003b} are able to fit individual components, they find on average lower spin temperatures for the CNM in the range of $\sim 40-70\,$K. Given that we can not distinguish between the CNM and the WNM, the measured spin temperature is a column density weighted mean between these two components \citep{Strasser2004}. This could explain the increased spin temperature reported in Sec. \ref{results_HI_spin_temp}. Another explanation for the increased spin temperature could be the strong radiation field of W43. \\
Further complications in our sample are the Galactic continuum sources. As these only trace the optical depth up to their location within the Milky Way, we miss the optical depth of the hydrogen behind the continuum source. On the other hand, the off-position measurement traces all the hydrogen along the line of sight. As the kinematic distance is uncertain due to near/far ambiguities, we cannot use the velocity to distinguish the distance of the emitting and absorbing hydrogen. Hence we underestimate the optical depth for the Galactic continuum sources and therefore overestimate the spin temperature. However, we do not see a significant difference in the mean spin temperature for Galactic and extragalactic continuum sources and we can neglect this effect.\\

\subsection{Column density and mass estimate}
\label{discussion_column_density}
In Sect. \ref{results_column_den_with_correction} we explained that we correct the \ion{H}{i} column density for optical depth effects as well as for the diffuse continuum emission. This leads to a more accurate estimate of the column density and a more accurate mass estimate. Nevertheless these corrections have limitations which we discuss in the following section.\\
As we need a strong continuum source in the background to determine the optical depth, we can measure the optical depth only toward certain locations. For our \ion{H}{i} column density correction, we used the strong continuum source W43-Main. Furthermore we assumed that the optical depth is the same for the entire cloud. This assumption might not hold, especially for the outer parts of the cloud. To investigate this effect, we compared the optical depth measurements for other sources at other positions. However, in the range of $v_{\rm{lsr}} = 80-110$\,km\,s$^{-1}$, we find that the optical depth is mostly saturated and determined by the corresponding lower limit. Two examples are given in the third panel of Fig. \ref{figure_continuum_extra_gal} and \ref{figure_continuum_gal}, which reveal a lower limit of $\tau_{\rm{lower-limit}} = 2.9$ and $\tau_{\rm{lower-limit}} = 1.9$, respectively. Other continuum sources that have a larger separation from the Galactic mid plane show similar results. For example the continuum source G30.699$-$0.630 has a Galactic latitude of $b \approx 0.6$\degr and still shows a saturated optical depth for the velocity range of W43 with a lower limit of $\tau_{\rm{lower-limit}} = 1.5$. If we use this continuum source to correct for optical depth effects and the weak diffuse continuum emission, we determine a mass of $M = 4.8 \times 10^6$\,M$_{\odot}$ for the whole cloud ($l = 29.0-31.5$\degr, $|b| \leq 1$\degr, $v_{\rm{lsr}} = 60-120$\,km\,s$^{-1}$). Hence, by using W43-Main to correct the optical depth for the entire cloud, we do not overestimate the mass in the outer parts significantly, but at most by a factor of 1.4. We will also use this mass estimate as a lower limit for the mass of W43.\\
The saturation of the optical depth is the second limitation we have to consider, especially for the inner part of the cloud. In this region we underestimate the opacity and, therefore, the column density which would lead to a further underestimate of the mass. Estimating this effect is difficult and therefore we report only lower limits for the column density and the mass in the inner part of W43.\\
The third limitation is the distance of the continuum source that we use to determine the optical depth. As explained in Sect. \ref{results_location_of_continuum_sources}, the continuum source W43-Main is located close to the tangential point of the Scutum-Centaurus arm at a distance of 5.5\,kpc. Hence, we only see \ion{H}{i} absorption as far as this distance, and miss all the \ion{H}{i} that is located behind the continuum source but still within the cloud. If we assume that the continuum source is at the center of the cloud, we underestimate the column density by a factor of two. Another approach to overcome this limitation is to look at other more distant continuum sources. The continuum source G31.388$-$0.384 is extragalactic and has a comparable brightness temperature to W43-Main. Hence, it is an ideal candidate for this test. We used the optical depth shown in the third panel of Fig. \ref{figure_continuum_extra_gal} to correct the \ion{H}{i} column density and measured the corrected mass for the same area and velocity range, as in Sect. \ref{results_column_den_with_correction}. The absoprtion spectrum of W43-Main and G31.388$-$0.384 are similar, except for the velocity range $v_{\rm{lsr}} = 100-120$\,km\,s$^{-1}$, which is missing in the W43-Main spectrum. We can also use the optical depth measurements of G31.388$-$0.384 to correct the \ion{H}{i} emission. However, as these absorption spectra are similar, the corrected masses are the same within the uncertainties ($M = 6.9 \times 10^6$\,M$_{\odot}$ for the correction with the optical depth of G31.388$-$0.384).\\
Summing up, we might overestimate the mass in the outer part of W43, but underestimate the mass in the inner part of W43. Due to the position of the continuum source, we might underestimate the mass by a factor of two. As explained in Sect. \ref{results_HISA}, we also miss some \ion{H}{i} due to self-absorption. Hence we report the \ion{H}{i} mass of W43 to be a lower limit of $M = 6.6_{-1.8} \times 10^6$\,M$_{\odot}$.\\
Several \ion{H}{i} mass estimates are given in the literature \citep{Nguyen2011,Motte2014}. All of them are calculated with the assumption of optically thin emission.\\
\citet{Motte2014} measured the \ion{H}{i} mass in the inner part of W43 ($l=29^{\circ}.6 \sim 31^{\circ}.4$ and $b=-0^{\circ}.5 \sim 0^{\circ}.3$) and for the velocity range ($v_{\rm{lsr}} = 60-120$\,km\,s$^{-1}$). The assumption of optically thin emission reveals an \ion{H}{i} mass of $M = 0.9 \times 10^6$\,M$_{\odot}$. If we use our corrected \ion{H}{i} column density to measure the mass in this area, we determine a \ion{H}{i} mass of $M = 2.2 \times 10^6$\,M$_{\odot}$ (this value is smaller than the value given in Sect. \ref{results_column_den_with_correction} as we only consider the inner part of W43). As outlined above, we claim that this mass estimate is a lower limit, and therefore the mass determined with the optically thin emission is at least a factor of 2.4 too small. This has implications for theoretical models, which we will discuss in Sect. \ref{discussion_column_density_threshold}.

\subsection{Spatial distribution of hydrogen}
\label{spatial_distribution_of_hydrogen}
\begin{figure}
  \resizebox{\hsize}{!}{\includegraphics{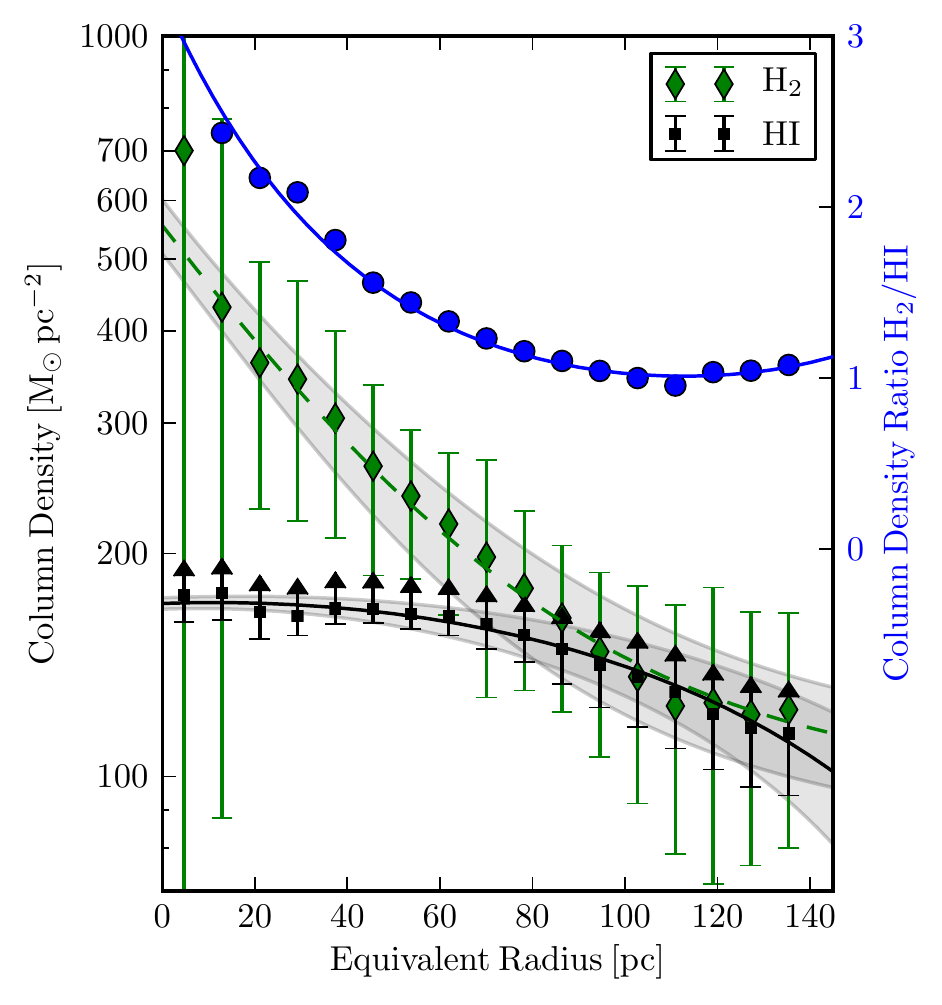}}
  \caption{The column density of \ion{H}{i} and H$_{\rm{2}}$ measured in elliptical annuli around W43 MM1. The x-axis presents the equivalent radius of these elliptical annuli. The black and green lines represent the fitted curves with the corresponding 1$\sigma$ uncertainties shown as a grey shaded area. The larger symbols (diamonds for H$_{\rm{2}}$, squares for \ion{H}{i}) present the averaged value of the elliptical annuli and their corresponding variations within the annuli. As the optical depth spectra saturates, we can only determine lower limits for the \ion{H}{i} column density. The blue dots and line show the H$_{\rm{2}}$/\ion{H}{i} column density ratio of the data and the fitted curves, respectively.}
  \label{figure_radial_distribution_of_column_density}
\end{figure}
\begin{figure}
  \resizebox{\hsize}{!}{\includegraphics{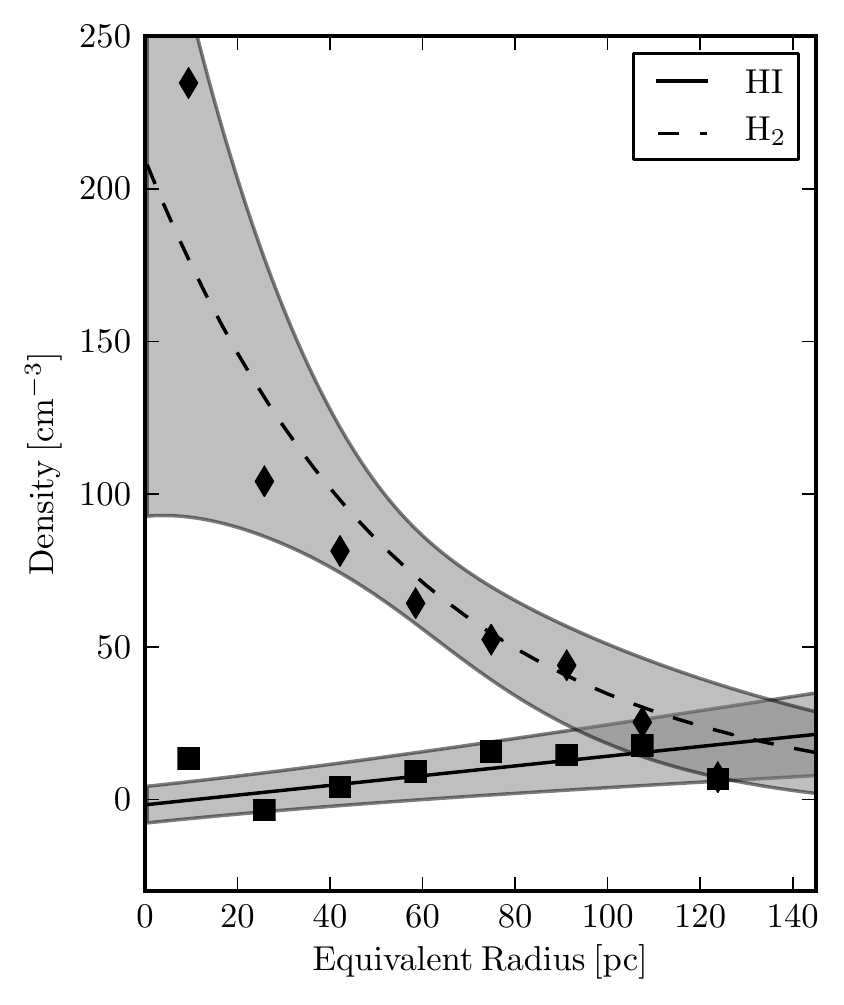}}
  \caption{The density of \ion{H}{i} and H$_{\rm{2}}$ as a function of the distance toward the center of W43 (equivalent radius). The diamonds and the dashed line represent the molecular hydrogen, whereas squares and the solid line represent the atomic hydrogen. The corresponding 3$\sigma$ uncertainties from the fit introduced in equation \ref{equation_density_of_HI_and_H2} are shown as grey shaded areas. }
  \label{figure_radial_distribution_of_density}
\end{figure}
Beside the total \ion{H}{i} mass, knowing the spatial distribution of the \ion{H}{i} is crucial to understand the formation of W43. As we have corrected the column density map for optical depth effects and the weak diffuse continuum emission, we can use this corrected column density map (see Fig. \ref{figure_HI_column_density_comparison}) to investigate the \ion{H}{i} spatial distribution in more detail, especially in the center.\\
Similar work was done by \citet{Motte2014}. They measured the \ion{H}{i} column density in rectangular annuli around the center of W43 ($l=30.5\degr$, $b = 0\degr$) with an aspect ratio of 3:2 and find an increasing \ion{H}{i} column density inwards from $N_{\rm{\ion{H}{i}}} \sim 40$\,M$_{\odot}$\,pc$^{-2}$ at a distance of 170\,pc to $N_{\rm{\ion{H}{i}}} \sim 80$\,M$_{\odot}$\,pc$^{-2}$ at a distance of 50\,pc from the center (velocity range $v_{\rm{lsr}} = 60-120$\,km\,s$^{-1}$). Since they assume optically thin \ion{H}{i} emission, they underestimate the \ion{H}{i} column density, especially in the central region (as discussed in Sect. \ref{discussion_column_density}).\\
To compensate for the approximate elliptical structure of W43, we choose elliptical annuli with an aspect ratio of 3:2 for major and minor axis, which fits the large scale structure of W43 well (see Fig. \ref{figure_HI_column_density_comparison}). As we focus on W43-Main, we choose the most massive submillimeter peak W43-MM1 \citep[$l=30.8175\degr$ and $b = -0.0571\degr$,][]{Motte2003} as the center for the ellipses. The width of each annulus is 10\,pc for the major axis and 6.6\,pc for the minor axis. For each elliptical annulus, we calculated the equivalent radius $ r = \sqrt{major \cdot minor} $ and assigned these values as the distance to the center shown in Fig. \ref{figure_radial_distribution_of_column_density}.\\ 
The black squares in Fig. \ref{figure_radial_distribution_of_column_density} represent the \ion{H}{i} column density mean value and the corresponding standard deviation of each elliptical annulus. We confirm the result of \citet{Motte2014}, that the \ion{H}{i} column density rises inwards. Our corrections allow us for the first time to study the central region of W43 ($r<50$\,pc) and, within the uncertainties, we report a flat column density distribution in this inner region. However, we mention that this flatness could also be due to the underestimation of the column density in the central part as the optical depth saturates and therefore the column density is a lower limit.\\
Furthermore, the diamond symbols in Fig. \ref{figure_radial_distribution_of_column_density} present the column density of the molecular hydrogen. The H$_{\rm{2}}$ distribution is centrally concentrated and the column density rises steeply toward the center, which is different from the \ion{H}{i} profile. The large uncertainties in the center are due to the clumpiness of the molecular hydrogen as the most prominent clumps, such as W43-MM1, are located in the first two elliptical annuli. Beside W43-Main, the second most prominent molecular clump is W43-South (see Fig. \ref{figure_HI_column_density_comparison}). However, in this analysis we focus on W43-Main and therefore, we choose W43-MM1 as the center for the ellipses. Furthermore, we masked W43-South for this analysis, as it would introduce a large uncertainty at an equivalent radius of $r \sim 50$\,pc.\\
We use a non-linear least square method ('curvefit' in the scipy package) to fit the mean values of the thin elliptical annuli with their corresponding uncertainties. For the \ion{H}{i} column density we assume a quadratic radial distribution:
\begin{equation}
N_{\rm{\ion{H}{i}}}(r) = a \cdot r^2 + b \cdot r + c ,
\label{equation_radial_distribution_of_HI}
\end{equation}
where $N_{\rm{\ion{H}{i}}}(r)$ describes the \ion{H}{i} column density and r describes the equivalent radius. The free parameters a, b and c have the fitted values of $a = -3.9 \pm 0.7 \times 10^{-3}$\,M$_{\odot}$\,pc$^{-4}$, $b = 0.089 \pm 0.099$\,M$_{\odot}$\,pc$^{-3}$ and $c = 171 \pm 3$\,M$_{\odot}$\,pc$^{-2}$. For the H$_{\rm{2}}$ distribution we assume an exponential function to fit the data:
\begin{equation}
N_{\rm{H_2}}(r) = d \cdot \exp(-e \cdot r) + f ,
\label{equation_radial_distribution_of_H2}
\end{equation}
where $N_{\rm{H_2}}(r)$ describes the H$_{\rm{2}}$ column density and $r$ describes the equivalent radius. The free parameters $d$, $e$ and $f$ have the fitted values of $d = 458 \pm 44$\,M$_{\odot}$\,pc$^{-2}$, $e = 0.022 \pm 0.004$\,pc$^{-1}$ and $f = 97 \pm 15$\,M$_{\odot}$\,pc$^{-2}$.\\
The black and green lines in Fig. \ref{figure_radial_distribution_of_column_density} present the column density fits and the grey shaded areas represent their corresponding uncertainties. Both, the \ion{H}{i} and H$_{\rm{2}}$ distributions are well fit by the assumed functions. As the uncertainties for the innermost part of the H$_{\rm{2}}$ distribution ($r<20$\,pc) are large, the fitted curve deviates from the data points and this area has to be treated cautiously. We also tested different functions to fit the data, such as functions with only two free parameters, or a Gaussian distribution for the \ion{H}{i} distribution, but the results were similar within the uncertainties. Finally, we chose the aforementioned functions as they could reproduce well the ratio of the H$_{\rm{2}}$ and \ion{H}{i} column density. This ratio is shown in blue in Fig. \ref{figure_radial_distribution_of_column_density}. The circles represent the ratio of the data points, whereas the solid line represents the ratio of the fitted curves. The ratio stays fairly constant at $N_{\rm{H_2}}/N_{\rm{\ion{H}{i}}} \approx 1$ for 140\,pc\,>\,$r$\,>\,60\,pc. For this region, the \ion{H}{i} column density also rises to its maximum value of $N_{\rm{\ion{H}{i}}} \approx 170$\,M$_{\odot}$\,pc$^{-2}$. Further inwards ($r<60$\,pc), the \ion{H}{i} column density stays constant at this maximum value, whereas the H$_{\rm{2}}$ column density rises sharply. Hence, the H$_{\rm{2}}$/\ion{H}{i} ratio also rises sharply to values above three. Summing up, the column density measurements imply that we have a mixture of \ion{H}{i} and H$_{\rm{2}}$ in the outskirt of the cloud and a molecular dominated region in the center. In the following, we will investigate this structure for the particle density.\\
As mentioned before, we find a flat column density for the \ion{H}{i} distribution toward the center of W43-Main, but what does this imply for the actual density in the center? If we assume that W43 has an elliptical shape, we can decompose the cloud into different layers, similar to an onion. Furthermore, we assign the appropriate column density to each layer with the information given in Fig. \ref{figure_radial_distribution_of_column_density}. As the column density is additive, the appropriate column density of each layer is the measured column density at the position of the considered layer minus the column density of all layers further outside. Hence, a flat \ion{H}{i} column density distribution toward the center (see Fig. \ref{figure_radial_distribution_of_column_density}) means that the layers in the center have no or a very small column density and therefore also a very small particle density. In the following section, we will use the concept of a elliptical layered structure to determine the actual particle density and show that indeed the \ion{H}{i} particle density drops toward the center of W43 within this model.

\subsection{Linking column density to particle density}
\label{discussion_linking_column_density_to_particle_density}

\begin{figure}
  \resizebox{\hsize}{!}{\includegraphics{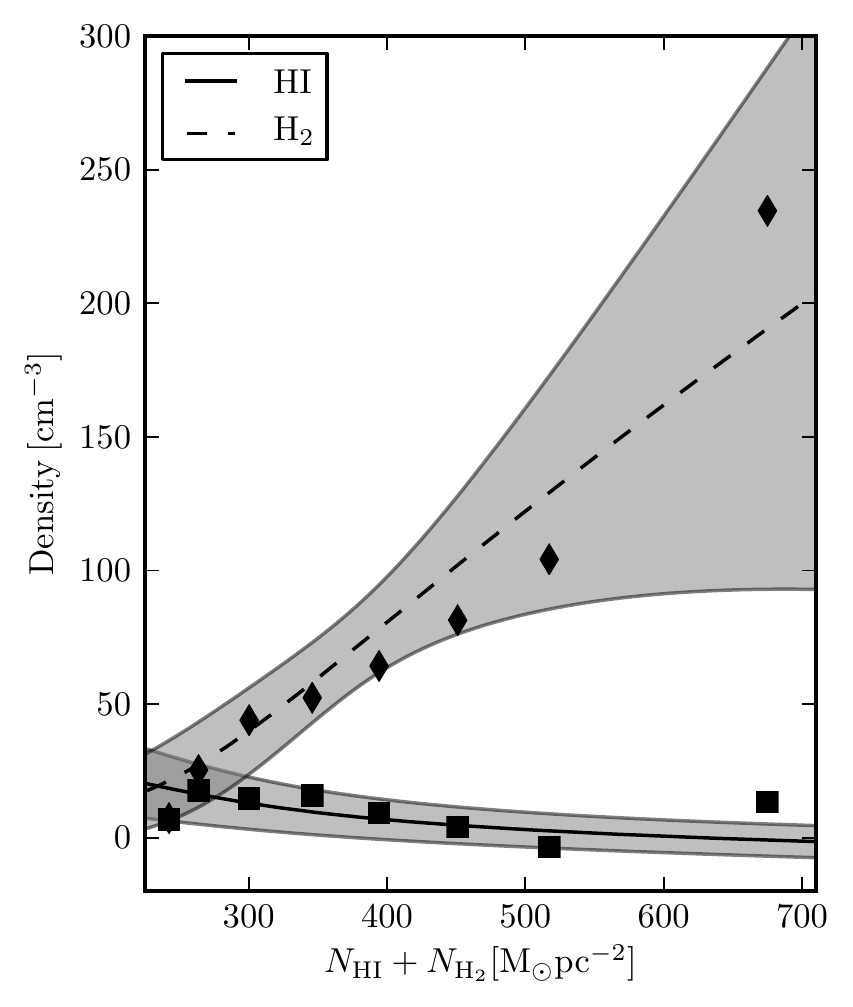}}
  \caption{The particle density of \ion{H}{i} and H$_{\rm{2}}$ as a function of the total column density (N$_{\rm{\ion{H}{i}}}$ + N$_{\rm{H_2}}$). The diamonds and the dashed line represents the molecular hydrogen, whereas squares the solid line represents the atomic hydrogen. The corresponding 3$\sigma$ uncertainties from the fit introduced in equations \ref{equation_radial_distribution_of_HI} and \ref{equation_radial_distribution_of_H2} are shown as grey shaded areas.}
  \label{figure_total_column_density-density_distribution}
\end{figure}

\begin{figure}
  \resizebox{\hsize}{!}{\includegraphics{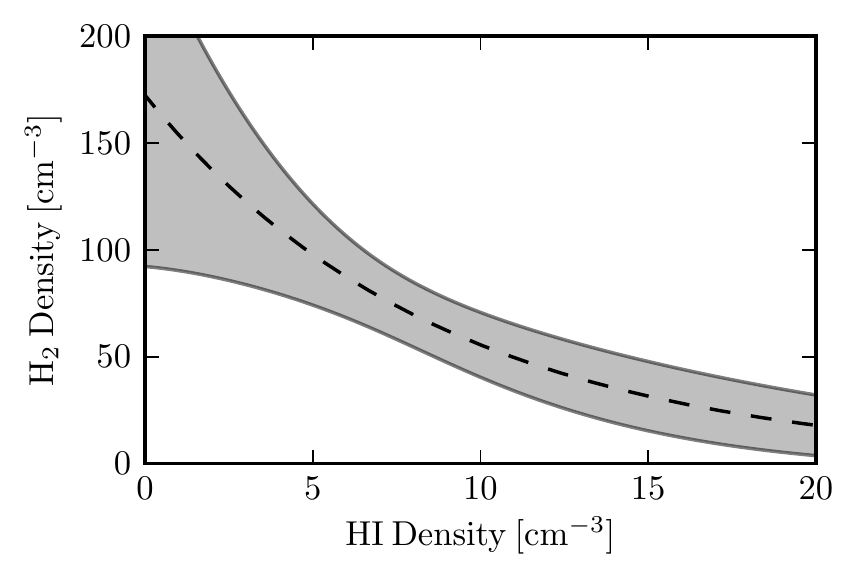}}
  \caption{The particle density of molecular hydrogen as a function of the atomic hydrogen. The black curve represents the best fitted curves from Fig. \ref{figure_radial_distribution_of_density} and the grey shaded area shows the 3$\sigma$ uncertainty for the H$_{\rm{2}}$ fit.}
  \label{figure_density-density_distribution}
\end{figure}

While the column density neglects the third dimension, it does not necessarily reflect the actual particle density. Hence, we have to estimate the third dimension. Using the elliptical, layered structure for W43 previously described and the results presented in Fig. \ref{figure_radial_distribution_of_column_density}, we estimate the particle density in each layer:
\begin{equation}
n_{\rm{i}} = \frac{\left(N_{\rm{i}} - N_{\rm{i+1}} \right)}{ 2\,\left(r_{\rm{i+1}}-r_{\rm{i}}\right) } ,
\label{equation_density_of_each_layer}
\end{equation}
where $n_{\rm{i}}$ describes the particle density of layer i, $N_{\rm{i}}$ and $N_{\rm{i+1}}$ describes the column density of layer i and layer i+1, respectively, and $r_{\rm{i}}$ and $r_{\rm{i+1}}$ describes the equivalent radius of layer i and i+1. The factor of two accounts for the two layers of the elliptical annuli in the front and the back of the cloud. The result for this calculation is shown in Fig. \ref{figure_radial_distribution_of_density} as squares and diamond data points for the \ion{H}{i} and H$_{\rm{2}}$ density, respectively. To increase the signal to noise ratio, we used larger elliptical annuli with a major axis of 20\,pc and a minor axis of 13.3\,pc.\\
In addition, we can use the fitted curves of the column density to estimate the density as well, by inserting equation \ref{equation_radial_distribution_of_HI} and \ref{equation_radial_distribution_of_H2} into equation \ref{equation_density_of_each_layer}. Hence, the radial distribution of the \ion{H}{i} and H$_{\rm{2}}$ densities are:
\begin{equation}
\begin{split}
&n_{\rm{\ion{H}{i}}} = -0.5\, (a \, \left(r_{\rm{i}}+r_{\rm{i+1}}\right) - b) , \\
&n_{\rm{H_2}} = \frac{d}{2 \, \left( r_{\rm{i+1}}-r_{\rm{i}} \right) } \,  \left( \exp(-e \cdot  r_{\rm{i}} ) - \exp(-e \cdot  r_{\rm{i+1}})  \right) ,
\end{split}
\label{equation_density_of_HI_and_H2}
\end{equation}
where $n_{\rm{\ion{H}{i}}}$ and $n_{\rm{H_2}}$ describes the particle density of \ion{H}{i} and H$_{\rm{2}}$, respectively, $r_{\rm{i}}$ describes the equivalent radius of layer i and $a$, $b$, $c$ and $d$ are the free fit parameters introduced in equations \ref{equation_radial_distribution_of_HI} and \ref{equation_radial_distribution_of_H2}. In Fig. \ref{figure_radial_distribution_of_density}, the calculated density distributions are shown as black lines with the corresponding uncertainties as a grey shaded areas. We note that the density distributions of \ion{H}{i} and H$_{\rm{2}}$ are very distinct. While the \ion{H}{i} distribution follows a simple linear relation, the H$_{\rm{2}}$ distribution shows an exponential increase toward the center. The particle density shows a mixture of the atomic and molecular hydrogen in the outskirt of W43 ($r>100$\,pc) and a molecular dominated interior, similar to the measurements of the column density. However, the particle density of the atomic hydrogen drops to almost zero toward the center which results in the observed constant column density.\\
Another way to present our results is shown in Fig. \ref{figure_total_column_density-density_distribution}. This figure shows the particle density of \ion{H}{i} and H$_{\rm{2}}$ introduced in equation \ref{equation_density_of_HI_and_H2} as a function of the total column density (N$_{\rm{\ion{H}{i}}}$ + N$_{\rm{H_2}}$). High column density regions, i.e. the central region of W43, are dominated by molecular hydrogen. On the other hand, we find an equivalent mixture of atomic and molecular hydrogen for low column density regions, which represent the envelope of W43.\\
In addition, we study the correlation between the \ion{H}{i} and H$_{\rm{2}}$ density. To do so, we replace the equivalent radius in Fig. \ref{figure_radial_distribution_of_density} and plot the H$_{\rm{2}}$ density as a function of the \ion{H}{i} density. Figure \ref{figure_density-density_distribution} shows the corresponding plot with the uncertainties for the H$_{\rm{2}}$ density. The statistical and systematical uncertainties for this plot are relatively large, but nevertheless we see an anti-correlation between the \ion{H}{i} and H$_{\rm{2}}$ density. This can be explained with a simple model that H$_{\rm{2}}$ forms out of \ion{H}{i} in the innermost part of W43. However, we do not detect a sharp transition between H$_{\rm{2}}$ and \ion{H}{i} predicted by \citet{Krumholz2008, Krumholz2009}. In the following section, we will discuss possible implications.

\subsection{Column density threshold for atomic hydrogen}
\label{discussion_column_density_threshold}
A fundamental question of molecular cloud formation is how does molecular hydrogen form out of neutral atomic hydrogen and what are the corresponding conditions \citep{Dobbs2014}. It is well known that the density must be high enough to shield the molecular hydrogen from the interstellar radiation field, to avoid dissociation back to its atomic form \citep{Hollenbach1997, McKee2007, Krumholz2008, Krumholz2009, MacLow2012}.\\ 
To describe this transition, \citet{Krumholz2008, Krumholz2009} suggest an analytic model, which is based on the assumption of a spherical cloud that is embedded in an isotropic external radiation field. They approximate that the transition between \ion{H}{i} in the envelope and H$_{\rm{2}}$ in the center occurs in an infinitely thin shell. An important results is that an \ion{H}{i} column density of $\sim 10$\,M$_{\odot}$\,pc$^{-2}$ is necessary to shield the molecular hydrogen from the interstellar radiation field. If the cloud reaches this critical \ion{H}{i} column density the formation of molecular hydrogen is efficient enough that most of the hydrogen goes into molecular form and the column density of \ion{H}{i} remains relatively constant at this level. Therefore we should not expect to observe \ion{H}{i} column densities larger than $\sim 10$\,M$_{\odot}$\,pc$^{-2}$, contradicting our results presented in Fig. \ref{figure_radial_distribution_of_column_density}. Furthermore, they show, that this \ion{H}{i} column density threshold is independent of the external radiation field, but has a weak dependence on the metallicity of the gas.\\
This model has three free parameters. First, the ratio of the measured cold neutral medium (CNM) density $n_{\rm{CNM}}$ to the minimal CNM density $n_{\rm{min}}$:  
\begin{equation}
n_{\rm{CNM}} = \phi_{\rm{CNM}} \cdot n_{\rm{min}}.
\label{equation_phi_CNM}
\end{equation}
The minimal CNM density is determined by the pressure balance with the warm neutral medium (WNM) and has a typical value of $n_{\rm{min}} \approx 7-8 \rm{cm^{-3}}$. As the range of pressure balance between the CNM and the WNM is limited, the maximum CNM density can be at most $\sim 10 \cdot n_{\rm{min}}$. Hence, $\phi_{\rm{CNM}}$ can vary between 1 and 10. \citet{Krumholz2009} assume $\phi_{\rm{CNM}} \approx 3$ for their fiducial value. The second free parameter is the ratio of the CNM density $n_{\rm{CNM}}$ to the molecular density $n_{\rm{mol}}$:
\begin{equation}
n_{\rm{mol}} = \phi_{\rm{mol}} \cdot n_{\rm{CNM}}.
\label{equation_phi_mol}
\end{equation}

\begin{table}
\caption{Densities and model parameter.}             
\label{table_krumholz_parameter}      
\centering    
\begin{tabular}{c c c c c c } 
\hline\hline
r & n$_{\rm{HI}}$ & n$_{\rm{H_2}}$ & $\phi_{\rm{mol}}$ & $\phi_{\rm{CNM}}$ & $\phi_{\rm{CNM}}$  \\
$\rm{[pc]}$ & [cm$\rm{^{-3}}$] & [cm$\rm{^{-3}}$] &  & $(G_0=1)$  & $(G_0=5)$ \\
\hline
140 & 20$\pm$4 & 17$\pm$5 & 0.85 & 2.7 & 0.58 \\
80 & 11$\pm$3 & 49$\pm$5 & 4.4 & 1.5 & 0.32\\
40 & 5$\pm$2 & 101$\pm$8 & 20.2 & 0.7 & 0.14\\
\hline   
\end{tabular} 
\tablefoot{\ion{H}{i} and H$_{\rm{2}}$ particle density for different equivalent radii with corresponding 1$\sigma$ uncertainty, extracted using the information given in Fig. \ref{figure_radial_distribution_of_density} and using equation \ref{equation_density_of_HI_and_H2}. The given uncertainties are the statistical uncertainties, but do not take into account the systematical uncertainties due to the saturation of the optical depth spectra. The model parameters $\phi_{\rm{mol}}$ and $\phi_{\rm{CNM}}$ are calculated using equations \ref{equation_phi_mol} and \ref{equation_phi_CNM}, respectively. $\phi_{\rm{CNM}}$ is calculated for solar values ($G_0 = 1$, $Z=1$) and for more realistic values of W43 ($G_0 = 5$, $Z=1.4$).}
\end{table}
\begin{figure}
  \resizebox{\hsize}{!}{\includegraphics{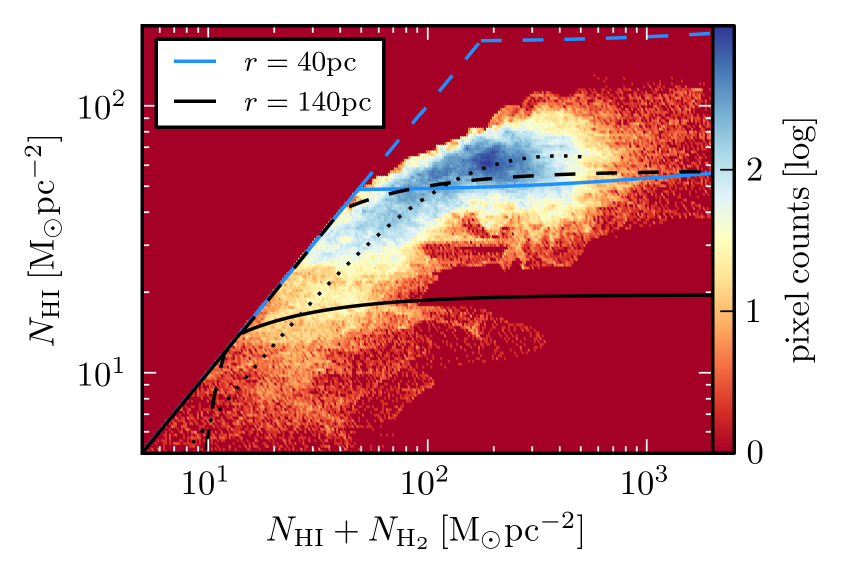}}
  \caption{The column density of the atomic hydrogen ($v_{\rm{lsr}} = 60-120$\,km\,s$^{-1}$) as a function of the total hydrogen column density (N$_{\rm{\ion{H}{i}}}$ + N$_{\rm{H_2}}$). The column density for the envelope ($r>140pc$, $N_{\rm{\ion{H}{i}}} =107\pm20$\,M$_{\odot}$\,pc$^{-2}$, $N_{\rm{H_2}} = 116\pm18$\,M$_{\odot}$\,pc$^{-2}$) is subtracted. The solid and dashed lines describe the theoretical model of \citet{Krumholz2008, Krumholz2009} for solar values ($G_0 = 1$, $Z=1$) and for more realistic values of W43 ($G_0 = 5$, $Z=1.4$), respectively. The blue and black colors represent the model parameters at different equivalent radii of $r=40$ and 140\,pc, respectively. The black dotted line shows the fitted curve of the elliptical annuli analysis introduced in Sect. \ref{discussion_column_density}.}
  \label{figure_krumholz-plot_column_density}
\end{figure}
\citet{Krumholz2009} suggests that this ratio is $\phi_{\rm{mol}} \approx 10$ and should not vary significantly. The last free parameter is the metallicity $Z$.\\
This model is supported by extragalactic observations \citep{Krumholz2009} as well as observations in the Perseus Molecular Cloud by \citet{Lee2012}. They find a uniform \ion{H}{i} column density of $N_{\rm{\ion{H}{i}}} \sim 6-8$\,M$_{\odot}$\,pc$^{-2}$ for H$_{\rm{2}}$ column densities up to $N_{\rm{H_2}} \sim 80$\,M$_{\odot}$\,pc$^{-2}$. To fit their data to the model of \citet{Krumholz2009}, they fixed $\phi_{\rm{mol}} = 10$ and $Z = 1.0 \, Z_{\odot}$ and fitted $\phi_{\rm{CNM}}$ which reveals values of $\phi_{\rm{CNM}} \approx 6-10$.\\
W43 is a more extreme test case for this theory as we have a much larger mass reservoir as well as an OB cluster that acts as a strong radiation source in the center. For this analysis, we extracted the column density of \ion{H}{i} and H$_{\rm{2}}$ for each pixel. As a basis for the \ion{H}{i} column density we chose the optical depth corrected version with a velocity range of $v_{\rm{lsr}} = 60-120$\,km\,s$^{-1}$. For the H$_{\rm{2}}$ column density, we again use the HiGal data (see Sect. \ref{observations_H2}). Even though the data base is the same as that used in Sect. \ref{discussion_linking_column_density_to_particle_density}, we stress that the method is different. Here we conduct a pixel by pixel comparison, whereas in Sect. \ref{discussion_linking_column_density_to_particle_density} we averaged the column density over elliptical annuli. In addition, we will focus merely on the inner part of W43 with $r<140$\,pc and therefore, we have to subtract the column density of the envelope. We assess the column density at $r = 140$\,pc as the envelope column density and subtract this value. The column density for the \ion{H}{i} and H$_{\rm{2}}$ envelopes are $N_{\rm{\ion{H}{i}}} =107\pm20$\,M$_{\odot}$\,pc$^{-2} = 1.3 \pm 0.3 \times 10^{22} \, \rm{cm^{-2}}$ and $N_{\rm{H_2}} = 116\pm18$\,M$_{\odot}$\,pc$^{-2} = 7.3 \pm 1.1 \times 10^{21} \, \rm{cm^{-2}}$, respectively.\\
Figure \ref{figure_krumholz-plot_column_density} shows the result of the pixel by pixel comparison of the \ion{H}{i} and H$_{\rm{2}}$ column density. Since we subtracted the column density of the envelope for this analysis, the values are smaller than in Fig. \ref{figure_radial_distribution_of_column_density}. For better readability, we do not show each single pixel comparison, but we performed a pixel binning. We do not observe the predicted threshold for the \ion{H}{i} column density of $\sim10$\,M$_{\odot}$\,pc$^{-2}$, instead our data show high \ion{H}{i} column density values, peaking between $N_{\rm{\ion{H}{i}}} =50-80$\,M$_{\odot}$\,pc$^{-2}$. Nevertheless we will try to fit the analytic model of \citet{Krumholz2009} to our data.\\
As described in Sec. \ref{discussion_linking_column_density_to_particle_density}, we use elliptical annuli to estimate the particle density. We can use this information (Fig. \ref{figure_radial_distribution_of_density}) to determine the model parameters $\phi_{\rm{CNM}}$ and $\phi_{\rm{mol}}$ using equations \ref{equation_phi_CNM} and \ref{equation_phi_mol}, respectively. Given that the \ion{H}{i} and H$_{\rm{2}}$ densities vary as a function of equivalent radius, the model parameters $\phi_{\rm{CNM}}$ and $\phi_{\rm{mol}}$ will also vary. We calculate the model parameters for three different distances $r=40$, 80 and 140\,pc that characterize the molecular dominated interior, the transition region and the well mixed outer area. To calculate $\phi_{\rm{CNM}}$ we have to know $n_{\rm{min}}$ which is given by \citet{Krumholz2009} as:
\begin{equation}
n_{\rm{min}} \approx 31\frac{G_0}{1+3.1\,Z^{0.365}} \rm{cm^{-3}},
\label{equation_phi_mol}
\end{equation}
where $G_0$ and $Z$ are the far-ultraviolet radiation intensity (in units of the Habing field) and the total metallicity, both normalized to their values in the solar neighborhood. For W43, we have a slightly higher metallicity of $Z=1.4Z_{\odot}$ \citep{Motte2014} and a large radiation field which could be up to $>100\,G_0$ \citep{Beuther2014} in the central region. We assume a more moderate radiation field for the outer regions of W43 with values around $5-10\,G_0$. In Table \ref{table_krumholz_parameter}, we present the \ion{H}{i} and H$_{\rm{2}}$ density as well as the model parameters for different distances. The parameter $\phi_{\rm{CNM}}$ is calculated for solar values ($G_0 = 1$, $Z=1$) and for more realistic values of W43 ($G_0 = 5$, $Z=1.4$) that results in $n_{\rm{min}} = 7.5 \rm{cm^{-3}}$ and $n_{\rm{min}} = 34.4 \rm{cm^{-3}}$, respectively. The solid blue and black lines in Fig. \ref{figure_krumholz-plot_column_density} show the theoretical column density for solar values and for the equivalent radius of $r=40$ and 140\,pc, respectively, whereas the dashed lines represent the more realistic values of W43 ($G_0 = 5$, $Z=1.4$). The dashed black line, which represents the outer area of W43 ($r=140$\,pc) with a moderate radiation field ($G_0 = 5$, $Z=1.4$) might fit the data. However, the model parameters might have unrealistically low values of $\phi_{\rm{CNM}} = 0.58$ and $\phi_{\rm{mol}}=0.85$. Using a stronger radiation field will increase $n_{\rm{min}}$ and therefore decrease $\phi_{\rm{CNM}}$ to even lower values. Hence, values with smaller distances and/or higher radiation fields predict column densities that are too high.\\
We do not have a conclusive answer why the model suggested by \citet{Krumholz2008, Krumholz2009} does not describe W43, but we suggest that the strong UV radiation produced by the central OB cluster \citep{Blum1999} and several further clusters in the environment are responsible for the dissociation of the molecular hydrogen. Another explanation was given by \citet{Motte2014}, They performed a similar analysis and their conclusion was that the analytical model by \citet{Krumholz2008, Krumholz2009} cannot describe a complicated molecular cloud complex, as we see several transition layers between \ion{H}{i} and H$_{\rm{2}}$ along a single line of sight and the assumption of a simple spherical cloud, without internal radiation sources breaks down.

\section{Conclusions}

The \ion{H}{i}, OH, Recombination Line Survey of the Milky Way (THOR) is a Galactic plane survey covering a large portion of the Galactic disk ($l=15-67\degr$, $|b| \leq 1$\degr). We use the VLA to observe the 21cm \ion{H}{i} line, 4 OH lines, 19 H$\alpha$ recombination lines and the continuum from 1-2\,GHz. In this paper we present the \ion{H}{i} data of the pilot field centered on the GMC associated with W43 ($l = 29.2-31.5$\degr, $|b| \leq 1$\degr). The main results can be summarized as:

\begin{enumerate}
\item
We measured the average spin temperature of the neutral hydrogen gas along the line of sight towards eight strong continuum sources. Half of them are Galactic and half of them are extragalactic. We find a median value of $T_{\rm{S}} = 95.7$\,K, which is in agreement with other studies.
\item
We can estimate the optical depth for the \ion{H}{i} line toward strong continuum sources at various locations in W43. The measured optical depth saturates for the main velocity component of W43 ($v_{\rm{lsr}} = 80-100$\,km\,s$^{-1}$) with lower limits of $\tau \sim 2.7$ in the center. Hence, the derived \ion{H}{i} masses based on optically thin emission strongly underestimates the hydrogen content. Employing further corrections for the weak and diffuse continuum emission, we obtain a lower limit for the \ion{H}{i} mass of $M = 6.6_{-1.8} \times 10^6$\,M$_{\odot}$ for a velocity range of $v_{\rm{lsr}} = 60-120$\,km\,s$^{-1}$ and an area of $l = 29.0-31.5$\degr and $|b| \leq 1$\degr. This is a factor of $\sim$2.4 larger than the \ion{H}{i} mass estimates with the assumption of optically thin emission.
\item
The measured \ion{H}{i} column density exceeds values of $N_{\rm{\ion{H}{i}}}\sim 150$\,M$_{\odot}$\,pc$^{-2}$ over much of the inner region with $r<80$\,pc. This is an order of magnitude larger than for low mass star forming regions such as Perseus. 
\item
As we corrected the \ion{H}{i} column density for optical depth effects and the weak continuum emission, we are able to study the \ion{H}{i} distribution spatially even in the innermost part of W43. We assumed an elliptical layered structure for the GMC associated with W43. This allows us to reconstruct the particle density of \ion{H}{i} and we find a linearly decreasing \ion{H}{i} density toward the center of W43 with values from $n_{\rm{\ion{H}{i}}} = 20$ to almost $0$\,cm$^{-3}$. Furthermore, we compared our results to the density of molecular hydrogen based on Herschel continuum data. The density of molecular hydrogen shows an exponential increase toward the center of W43 with values rising from $n_{\rm{H_2}} = 15$ to $200$\,cm$^{-3}$. For smaller clumps, the density of H$_{\rm{2}}$ can rise to even higher values.
\item
We compared our measurements to the analytic model suggested by \citet{Krumholz2008,Krumholz2009}. Our data does not show a sharp transition between \ion{H}{i} and H$_{\rm{2}}$, neither do we find the predicted threshold for the \ion{H}{i} column density of $\sim 10$\,M$_{\odot}$\,pc$^{-2}$. Based on these models, larger \ion{H}{i} column densities should not exist, as molecular hydrogen should form for such high \ion{H}{i} column densities. To fit the model, we have to assume low model parameters, which may indicate that the model is not applicable in a region with such a high radiation field. We suggest that the addition of an internal radiation field from a central cluster may be required to describe the observations. Thus, this work has shown that the \ion{H}{i} content of W43 and its relation to H$_{\rm{2}}$ challenges current models of H$_{\rm{2}}$ formation.
\end{enumerate}

\begin{acknowledgements}
We like to thank the unknown referee for an excellent report with a lot of helpful comments that improved the paper.\\
The National Radio Astronomy Observatory is a facility of the National Science Foundation operated under cooperative agreement by Associated Universities, Inc.\\
RSK, SCOG, RJS and SER acknowledge financial support by the Deutsche Forschungsgemeinschaft (DFG) via the Heidelberg Sonderforschungsbereich SFB 881 “The Milky Way System” (subprojects B1, B2, and B8) as well as the Schwerpunktprogramm SPP 1573 "Physics of the Interstellar Medium". RSK furthermore acknowledges support from the European Research Council under the European Community’s Seventh Framework Programme (FP7/2007-2013) via the ERC Advanced Grant STARLIGHT (project number 339177).\\
We gratefully acknowledge the help from the team of the Pete V. Domenici Science Operations Center (SOC) in Socorro for their help on the data reduction during an extended visit in 2013.\\
S. B. is a fellow of the International Max Planck Research School for Astronomy and Cosmic Physics (IMPRS) at the University of Heidelberg and acknowledges support.
\end{acknowledgements}

\bibliographystyle{aa}
\bibliography{../references.bib}

\begin{thebibliography}{66}
\expandafter\ifx\csname natexlab\endcsname\relax\def\natexlab#1{#1}\fi

\bibitem[{{Anderson} {et~al.}(2014){Anderson}, {Bania}, {Balser}, {Cunningham},
  {Wenger}, {Johnstone}, \& {Armentrout}}]{Anderson2014}
{Anderson}, L.~D., {Bania}, T.~M., {Balser}, D.~S., {et~al.} 2014, \apjs, 212,
  1

\bibitem[{{Anderson} {et~al.}(2011){Anderson}, {Bania}, {Balser}, \&
  {Rood}}]{Anderson2011}
{Anderson}, L.~D., {Bania}, T.~M., {Balser}, D.~S., \& {Rood}, R.~T. 2011,
  \apjs, 194, 32

\bibitem[{{Bally} {et~al.}(2010){Bally}, {Anderson}, {Battersby}, {Calzoletti},
  {Digiorgio}, {Faustini}, {Ginsburg}, {Li}, {Nguyen Luong}, {Molinari},
  {Motte}, {Pestalozzi}, {Plume}, {Rodon}, {Schilke}, {Schlingman},
  {Schneider-Bontemps}, {Shirley}, {Stringfellow}, {Testi}, {Traficante},
  {Veneziani}, \& {Zavagno}}]{Bally2010}
{Bally}, J., {Anderson}, L.~D., {Battersby}, C., {et~al.} 2010, \aap, 518, L90

\bibitem[{{Beuther} {et~al.}(2014){Beuther}, {Ragan}, {Ossenkopf}, {Glover},
  {Henning}, {Linz}, {Nielbock}, {Krause}, {Stutzki}, {Schilke}, \&
  {G{\"u}sten}}]{Beuther2014}
{Beuther}, H., {Ragan}, S.~E., {Ossenkopf}, V., {et~al.} 2014, \aap, 571, A53

\bibitem[{{Beuther} {et~al.}(2012){Beuther}, {Tackenberg}, {Linz}, {Henning},
  {Krause}, {Ragan}, {Nielbock}, {Launhardt}, {Schmiedeke}, {Schuller},
  {Carlhoff}, {Nguyen-Luong}, \& {Sakai}}]{Beuther2012}
{Beuther}, H., {Tackenberg}, J., {Linz}, H., {et~al.} 2012, \aap, 538, A11

\bibitem[{{Bhatnagar} {et~al.}(2013){Bhatnagar}, {Rau}, \&
  {Golap}}]{Bhatnagar2013}
{Bhatnagar}, S., {Rau}, U., \& {Golap}, K. 2013, \apj, 770, 91

\bibitem[{{Bhatnagar} {et~al.}(2011){Bhatnagar}, {Rau}, {Green}, \&
  {Rupen}}]{Bhatnagar2011}
{Bhatnagar}, S., {Rau}, U., {Green}, D.~A., \& {Rupen}, M.~P. 2011, \apjl, 739,
  L20

\bibitem[{{Blum} {et~al.}(1999){Blum}, {Damineli}, \& {Conti}}]{Blum1999}
{Blum}, R.~D., {Damineli}, A., \& {Conti}, P.~S. 1999, \aj, 117, 1392

\bibitem[{{Brunthaler} {et~al.}(2011){Brunthaler}, {Reid}, {Menten}, {Zheng},
  {Bartkiewicz}, {Choi}, {Dame}, {Hachisuka}, {Immer}, {Moellenbrock},
  {Moscadelli}, {Rygl}, {Sanna}, {Sato}, {Wu}, {Xu}, \&
  {Zhang}}]{Brunthaler2011}
{Brunthaler}, A., {Reid}, M.~J., {Menten}, K.~M., {et~al.} 2011, Astronomische
  Nachrichten, 332, 461

\bibitem[{{Carlhoff} {et~al.}(2013){Carlhoff}, {Nguyen Luong}, {Schilke},
  {Motte}, {Schneider}, {Beuther}, {Bontemps}, {Heitsch}, {Hill}, {Kramer},
  {Ossenkopf}, {Schuller}, {Simon}, \& {Wyrowski}}]{Carlhoff2013}
{Carlhoff}, P., {Nguyen Luong}, Q., {Schilke}, P., {et~al.} 2013, \aap, 560,
  A24

\bibitem[{{Chengalur} {et~al.}(2013){Chengalur}, {Kanekar}, \&
  {Roy}}]{Chengalur2013}
{Chengalur}, J.~N., {Kanekar}, N., \& {Roy}, N. 2013, \mnras, 432, 3074

\bibitem[{{Clark}(1965)}]{Clark1965}
{Clark}, B.~G. 1965, \apj, 142, 1398

\bibitem[{{Dobbs} {et~al.}(2014){Dobbs}, {Krumholz}, {Ballesteros-Paredes},
  {Bolatto}, {Fukui}, {Heyer}, {Low}, {Ostriker}, \&
  {V{\'a}zquez-Semadeni}}]{Dobbs2014}
{Dobbs}, C.~L., {Krumholz}, M.~R., {Ballesteros-Paredes}, J., {et~al.} 2014,
  Protostars and Planets VI, 3

\bibitem[{{Fukui} {et~al.}(2015){Fukui}, {Torii}, {Onishi}, {Yamamoto},
  {Okamoto}, {Hayakawa}, {Tachihara}, \& {Sano}}]{Fukui2015}
{Fukui}, Y., {Torii}, K., {Onishi}, T., {et~al.} 2015, \apj, 798, 6

\bibitem[{{Gibson} {et~al.}(2005{\natexlab{a}}){Gibson}, {Taylor}, {Higgs},
  {Brunt}, \& {Dewdney}}]{Gibson2005}
{Gibson}, S.~J., {Taylor}, A.~R., {Higgs}, L.~A., {Brunt}, C.~M., \& {Dewdney},
  P.~E. 2005{\natexlab{a}}, \apj, 626, 195

\bibitem[{{Gibson} {et~al.}(2005{\natexlab{b}}){Gibson}, {Taylor}, {Higgs},
  {Brunt}, \& {Dewdney}}]{Gibson2005b}
{Gibson}, S.~J., {Taylor}, A.~R., {Higgs}, L.~A., {Brunt}, C.~M., \& {Dewdney},
  P.~E. 2005{\natexlab{b}}, \apj, 626, 214

\bibitem[{{Gibson} {et~al.}(2000){Gibson}, {Taylor}, {Higgs}, \&
  {Dewdney}}]{Gibson2000}
{Gibson}, S.~J., {Taylor}, A.~R., {Higgs}, L.~A., \& {Dewdney}, P.~E. 2000,
  \apj, 540, 851

\bibitem[{{Glover} \& {Clark}(2012)}]{Glover2012}
{Glover}, S.~C.~O. \& {Clark}, P.~C. 2012, \mnras, 421, 9

\bibitem[{{Goldsmith} \& {Li}(2005)}]{Goldsmith2005}
{Goldsmith}, P.~F. \& {Li}, D. 2005, \apj, 622, 938

\bibitem[{{Heiles} \& {Troland}(2003{\natexlab{a}})}]{Heiles2003a}
{Heiles}, C. \& {Troland}, T.~H. 2003{\natexlab{a}}, \apjs, 145, 329

\bibitem[{{Heiles} \& {Troland}(2003{\natexlab{b}})}]{Heiles2003b}
{Heiles}, C. \& {Troland}, T.~H. 2003{\natexlab{b}}, \apj, 586, 1067

\bibitem[{{Hill} {et~al.}(2011){Hill}, {Motte}, {Didelon}, {Bontemps},
  {Minier}, {Hennemann}, {Schneider}, {Andr{\'e}}, {Men'shchikov}, {Anderson},
  {Arzoumanian}, {Bernard}, {di Francesco}, {Elia}, {Giannini}, {Griffin},
  {K{\"o}nyves}, {Kirk}, {Marston}, {Martin}, {Molinari}, {Nguyen Luong},
  {Peretto}, {Pezzuto}, {Roussel}, {Sauvage}, {Sousbie}, {Testi},
  {Ward-Thompson}, {White}, {Wilson}, \& {Zavagno}}]{Hill2011}
{Hill}, T., {Motte}, F., {Didelon}, P., {et~al.} 2011, \aap, 533, A94

\bibitem[{{Hoare} {et~al.}(2012){Hoare}, {Purcell}, {Churchwell}, {Diamond},
  {Cotton}, {Chandler}, {Smethurst}, {Kurtz}, {Mundy}, {Dougherty}, {Fender},
  {Fuller}, {Jackson}, {Garrington}, {Gledhill}, {Goldsmith}, {Lumsden},
  {Mart{\'{\i}}}, {Moore}, {Muxlow}, {Oudmaijer}, {Pandian}, {Paredes},
  {Shepherd}, {Spencer}, {Thompson}, {Umana}, {Urquhart}, \&
  {Zijlstra}}]{Hoare2012}
{Hoare}, M.~G., {Purcell}, C.~R., {Churchwell}, E.~B., {et~al.} 2012, \pasp,
  124, 939

\bibitem[{{Hollenbach} \& {Tielens}(1997)}]{Hollenbach1997}
{Hollenbach}, D.~J. \& {Tielens}, A.~G.~G.~M. 1997, \araa, 35, 179

\bibitem[{{Kainulainen} {et~al.}(2013){Kainulainen}, {Ragan}, {Henning}, \&
  {Stutz}}]{Kainulainen2013b}
{Kainulainen}, J., {Ragan}, S.~E., {Henning}, T., \& {Stutz}, A. 2013, \aap,
  557, A120

\bibitem[{{Kainulainen} \& {Tan}(2013)}]{Kainulainen2013}
{Kainulainen}, J. \& {Tan}, J.~C. 2013, \aap, 549, A53

\bibitem[{{Kalberla} \& {Kerp}(2009)}]{Kalberla2009}
{Kalberla}, P.~M.~W. \& {Kerp}, J. 2009, \araa, 47, 27

\bibitem[{{Krumholz} {et~al.}(2008){Krumholz}, {McKee}, \&
  {Tumlinson}}]{Krumholz2008}
{Krumholz}, M.~R., {McKee}, C.~F., \& {Tumlinson}, J. 2008, \apj, 689, 865

\bibitem[{{Krumholz} {et~al.}(2009){Krumholz}, {McKee}, \&
  {Tumlinson}}]{Krumholz2009}
{Krumholz}, M.~R., {McKee}, C.~F., \& {Tumlinson}, J. 2009, \apj, 693, 216

\bibitem[{{Kr{\v c}o} \& {Goldsmith}(2010)}]{Krco2010}
{Kr{\v c}o}, M. \& {Goldsmith}, P.~F. 2010, \apj, 724, 1402

\bibitem[{{Lada} {et~al.}(2007){Lada}, {Alves}, \& {Lombardi}}]{Lada2007}
{Lada}, C.~J., {Alves}, J.~F., \& {Lombardi}, M. 2007, Protostars and Planets
  V, 3

\bibitem[{{Lee} {et~al.}(2012){Lee}, {Stanimirovi{\'c}}, {Douglas}, {Knee}, {Di
  Francesco}, {Gibson}, {Begum}, {Grcevich}, {Heiles}, {Korpela}, {Leroy},
  {Peek}, {Pingel}, {Putman}, \& {Saul}}]{Lee2012}
{Lee}, M.-Y., {Stanimirovi{\'c}}, S., {Douglas}, K.~A., {et~al.} 2012, \apj,
  748, 75

\bibitem[{{Lester} {et~al.}(1985){Lester}, {Dinerstein}, {Werner}, {Harvey},
  {Evans}, \& {Brown}}]{Lester1985}
{Lester}, D.~F., {Dinerstein}, H.~L., {Werner}, M.~W., {et~al.} 1985, \apj,
  296, 565

\bibitem[{{Li} \& {Goldsmith}(2003)}]{Li2003}
{Li}, D. \& {Goldsmith}, P.~F. 2003, \apj, 585, 823

\bibitem[{{Liszt} {et~al.}(1993){Liszt}, {Braun}, \& {Greisen}}]{Liszt1993}
{Liszt}, H.~S., {Braun}, R., \& {Greisen}, E.~W. 1993, \aj, 106, 2349

\bibitem[{{Louvet} {et~al.}(2014){Louvet}, {Motte}, {Hennebelle}, {Maury},
  {Bonnell}, {Bontemps}, {Gusdorf}, {Hill}, {Gueth}, {Peretto},
  {Duarte-Cabral}, {Stephan}, {Schilke}, {Csengeri}, {Nguyen Luong}, \&
  {Lis}}]{Louvet2014}
{Louvet}, F., {Motte}, F., {Hennebelle}, P., {et~al.} 2014, \aap, 570, A15

\bibitem[{{Mac Low} \& {Glover}(2012)}]{MacLow2012}
{Mac Low}, M.-M. \& {Glover}, S.~C.~O. 2012, \apj, 746, 135

\bibitem[{{Mac Low} \& {Klessen}(2004)}]{MacLow2004}
{Mac Low}, M.-M. \& {Klessen}, R.~S. 2004, Reviews of Modern Physics, 76, 125

\bibitem[{{McClure-Griffiths} {et~al.}(2012){McClure-Griffiths}, {Dickey},
  {Gaensler}, {Green}, {Green}, \& {Haverkorn}}]{McClure-Griffiths2012}
{McClure-Griffiths}, N.~M., {Dickey}, J.~M., {Gaensler}, B.~M., {et~al.} 2012,
  \apjs, 199, 12

\bibitem[{{McClure-Griffiths} {et~al.}(2006){McClure-Griffiths}, {Dickey},
  {Gaensler}, {Green}, \& {Haverkorn}}]{McClure-Griffiths2006}
{McClure-Griffiths}, N.~M., {Dickey}, J.~M., {Gaensler}, B.~M., {Green}, A.~J.,
  \& {Haverkorn}, M. 2006, \apj, 652, 1339

\bibitem[{{McKee} \& {Ostriker}(2007)}]{McKee2007}
{McKee}, C.~F. \& {Ostriker}, E.~C. 2007, \araa, 45, 565

\bibitem[{{Molinari} {et~al.}(2010){Molinari}, {Swinyard}, {Bally}, {Barlow},
  {Bernard}, {Martin}, {Moore}, {Noriega-Crespo}, {Plume}, {Testi}, {Zavagno},
  {Abergel}, {Ali}, {Anderson}, {Andr{\'e}}, {Baluteau}, {Battersby},
  {Beltr{\'a}n}, {Benedettini}, {Billot}, {Blommaert}, {Bontemps}, {Boulanger},
  {Brand}, {Brunt}, {Burton}, {Calzoletti}, {Carey}, {Caselli}, {Cesaroni},
  {Cernicharo}, {Chakrabarti}, {Chrysostomou}, {Cohen}, {Compiegne}, {de
  Bernardis}, {de Gasperis}, {di Giorgio}, {Elia}, {Faustini}, {Flagey},
  {Fukui}, {Fuller}, {Ganga}, {Garcia-Lario}, {Glenn}, {Goldsmith}, {Griffin},
  {Hoare}, {Huang}, {Ikhenaode}, {Joblin}, {Joncas}, {Juvela}, {Kirk},
  {Lagache}, {Li}, {Lim}, {Lord}, {Marengo}, {Marshall}, {Masi}, {Massi},
  {Matsuura}, {Minier}, {Miville-Desch{\^e}nes}, {Montier}, {Morgan}, {Motte},
  {Mottram}, {M{\"u}ller}, {Natoli}, {Neves}, {Olmi}, {Paladini}, {Paradis},
  {Parsons}, {Peretto}, {Pestalozzi}, {Pezzuto}, {Piacentini}, {Piazzo},
  {Polychroni}, {Pomar{\`e}s}, {Popescu}, {Reach}, {Ristorcelli}, {Robitaille},
  {Robitaille}, {Rod{\'o}n}, {Roy}, {Royer}, {Russeil}, {Saraceno}, {Sauvage},
  {Schilke}, {Schisano}, {Schneider}, {Schuller}, {Schulz}, {Sibthorpe},
  {Smith}, {Smith}, {Spinoglio}, {Stamatellos}, {Strafella}, {Stringfellow},
  {Sturm}, {Taylor}, {Thompson}, {Traficante}, {Tuffs}, {Umana}, {Valenziano},
  {Vavrek}, {Veneziani}, {Viti}, {Waelkens}, {Ward-Thompson}, {White},
  {Wilcock}, {Wyrowski}, {Yorke}, \& {Zhang}}]{Molinari2010}
{Molinari}, S., {Swinyard}, B., {Bally}, J., {et~al.} 2010, \aap, 518, L100

\bibitem[{{Motte} {et~al.}(2014){Motte}, {Nguy{\^e}n Luong}, {Schneider},
  {Heitsch}, {Glover}, {Carlhoff}, {Hill}, {Bontemps}, {Schilke}, {Louvet},
  {Hennemann}, {Didelon}, \& {Beuther}}]{Motte2014}
{Motte}, F., {Nguy{\^e}n Luong}, Q., {Schneider}, N., {et~al.} 2014, \aap, 571,
  A32

\bibitem[{{Motte} {et~al.}(2003){Motte}, {Schilke}, \& {Lis}}]{Motte2003}
{Motte}, F., {Schilke}, P., \& {Lis}, D.~C. 2003, \apj, 582, 277

\bibitem[{{Murray} {et~al.}(2014){Murray}, {Lindner}, {Stanimirovi{\'c}},
  {Goss}, {Heiles}, {Dickey}, {Pingel}, {Lawrence}, {Jencson}, {Babler}, \&
  {Hennebelle}}]{Murray2014}
{Murray}, C.~E., {Lindner}, R.~R., {Stanimirovi{\'c}}, S., {et~al.} 2014,
  \apjl, 781, L41

\bibitem[{{Nguyen-Lu'o'ng} {et~al.}(2013){Nguyen-Lu'o'ng}, {Motte}, {Carlhoff},
  {Louvet}, {Lesaffre}, {Schilke}, {Hill}, {Hennemann}, {Gusdorf}, {Didelon},
  {Schneider}, {Bontemps}, {Duarte-Cabral}, {Menten}, {Martin}, {Wyrowski},
  {Bendo}, {Roussel}, {Bernard}, {Bronfman}, {Henning}, {Kramer}, \&
  {Heitsch}}]{Nguyen2013}
{Nguyen-Lu'o'ng}, Q., {Motte}, F., {Carlhoff}, P., {et~al.} 2013, \apj, 775, 88

\bibitem[{{Nguyen Luong} {et~al.}(2011){Nguyen Luong}, {Motte}, {Schuller},
  {Schneider}, {Bontemps}, {Schilke}, {Menten}, {Heitsch}, {Wyrowski},
  {Carlhoff}, {Bronfman}, \& {Henning}}]{Nguyen2011}
{Nguyen Luong}, Q., {Motte}, F., {Schuller}, F., {et~al.} 2011, \aap, 529, A41

\bibitem[{{Offner} {et~al.}(2014){Offner}, {Clark}, {Hennebelle}, {Bastian},
  {Bate}, {Hopkins}, {Moraux}, \& {Whitworth}}]{Offner2014}
{Offner}, S.~S.~R., {Clark}, P.~C., {Hennebelle}, P., {et~al.} 2014, Protostars
  and Planets VI, 53

\bibitem[{{Pineda} {et~al.}(2013){Pineda}, {Langer}, {Velusamy}, \&
  {Goldsmith}}]{Pineda2013}
{Pineda}, J.~L., {Langer}, W.~D., {Velusamy}, T., \& {Goldsmith}, P.~F. 2013,
  \aap, 554, A103

\bibitem[{{Purcell} {et~al.}(2013){Purcell}, {Hoare}, {Cotton}, {Lumsden},
  {Urquhart}, {Chandler}, {Churchwell}, {Diamond}, {Dougherty}, {Fender},
  {Fuller}, {Garrington}, {Gledhill}, {Goldsmith}, {Hindson}, {Jackson},
  {Kurtz}, {Mart{\'{\i}}}, {Moore}, {Mundy}, {Muxlow}, {Oudmaijer}, {Pandian},
  {Paredes}, {Shepherd}, {Smethurst}, {Spencer}, {Thompson}, {Umana}, \&
  {Zijlstra}}]{Purcell2013}
{Purcell}, C.~R., {Hoare}, M.~G., {Cotton}, W.~D., {et~al.} 2013, \apjs, 205, 1

\bibitem[{{Radhakrishnan} {et~al.}(1972){Radhakrishnan}, {Murray}, {Lockhart},
  \& {Whittle}}]{Radhakrishnan1972}
{Radhakrishnan}, V., {Murray}, J.~D., {Lockhart}, P., \& {Whittle}, R.~P.~J.
  1972, \apjs, 24, 15

\bibitem[{{Rau} {et~al.}(2014){Rau}, {Bhatnagar}, \& {Owen}}]{Rau2014}
{Rau}, U., {Bhatnagar}, S., \& {Owen}, F.~N. 2014, ArXiv e-prints

\bibitem[{{Reid} {et~al.}(2014){Reid}, {Menten}, {Brunthaler}, {Zheng}, {Dame},
  {Xu}, {Wu}, {Zhang}, {Sanna}, {Sato}, {Hachisuka}, {Choi}, {Immer},
  {Moscadelli}, {Rygl}, \& {Bartkiewicz}}]{Reid2014}
{Reid}, M.~J., {Menten}, K.~M., {Brunthaler}, A., {et~al.} 2014, \apj, 783, 130

\bibitem[{{Roy} {et~al.}(2013{\natexlab{a}}){Roy}, {Kanekar}, {Braun}, \&
  {Chengalur}}]{Roy2013}
{Roy}, N., {Kanekar}, N., {Braun}, R., \& {Chengalur}, J.~N.
  2013{\natexlab{a}}, \mnras, 436, 2352

\bibitem[{{Roy} {et~al.}(2013{\natexlab{b}}){Roy}, {Kanekar}, \&
  {Chengalur}}]{Roy2013b}
{Roy}, N., {Kanekar}, N., \& {Chengalur}, J.~N. 2013{\natexlab{b}}, \mnras,
  436, 2366

\bibitem[{{Smith} {et~al.}(2014){Smith}, {Glover}, {Clark}, {Klessen}, \&
  {Springel}}]{Smith2014}
{Smith}, R.~J., {Glover}, S.~C.~O., {Clark}, P.~C., {Klessen}, R.~S., \&
  {Springel}, V. 2014, \mnras, 441, 1628

\bibitem[{Stahler {et~al.}(2005)Stahler, Palla, \& Palla}]{Stahler2004}
Stahler, S.~W., Palla, F., \& Palla, F. 2005, The Formation of Stars (Physics
  Textbook), 1st edn. (Wiley-VCH)

\bibitem[{{Stil} {et~al.}(2006){Stil}, {Taylor}, {Dickey}, {Kavars}, {Martin},
  {Rothwell}, {Boothroyd}, {Lockman}, \& {McClure-Griffiths}}]{Stil2006}
{Stil}, J.~M., {Taylor}, A.~R., {Dickey}, J.~M., {et~al.} 2006, \aj, 132, 1158

\bibitem[{{Strasser} \& {Taylor}(2004)}]{Strasser2004}
{Strasser}, S. \& {Taylor}, A.~R. 2004, \apj, 603, 560

\bibitem[{{Strasser} {et~al.}(2007){Strasser}, {Dickey}, {Taylor}, {Boothroyd},
  {Gaensler}, {Green}, {Kavars}, {Lockman}, {Martin}, {McClure-Griffiths},
  {Rothwell}, \& {Stil}}]{Strasser2007}
{Strasser}, S.~T., {Dickey}, J.~M., {Taylor}, A.~R., {et~al.} 2007, \aj, 134,
  2252

\bibitem[{{Taylor} {et~al.}(2003){Taylor}, {Gibson}, {Peracaula}, {Martin},
  {Landecker}, {Brunt}, {Dewdney}, {Dougherty}, {Gray}, {Higgs}, {Kerton},
  {Knee}, {Kothes}, {Purton}, {Uyaniker}, {Wallace}, {Willis}, \&
  {Durand}}]{Taylor2003}
{Taylor}, A.~R., {Gibson}, S.~J., {Peracaula}, M., {et~al.} 2003, \aj, 125,
  3145

\bibitem[{{Vall{\'e}e}(2008)}]{Vallee2008}
{Vall{\'e}e}, J.~P. 2008, \aj, 135, 1301

\bibitem[{Wilson {et~al.}(2010)Wilson, Rohlfs, \&
  H{\"u}ttemeister}]{Wilson2010}
Wilson, T.~L., Rohlfs, K., \& H{\"u}ttemeister, S. 2010, Tools of Radio
  Astronomy (Astronomy and Astrophysics Library), softcover reprint of
  hardcover 5th ed. 2009 edn. (Springer)

\bibitem[{{Wolfire} {et~al.}(1995){Wolfire}, {Hollenbach}, {McKee}, {Tielens},
  \& {Bakes}}]{Wolfire1995}
{Wolfire}, M.~G., {Hollenbach}, D., {McKee}, C.~F., {Tielens}, A.~G.~G.~M., \&
  {Bakes}, E.~L.~O. 1995, \apj, 443, 152

\bibitem[{{Wolfire} {et~al.}(2003){Wolfire}, {McKee}, {Hollenbach}, \&
  {Tielens}}]{Wolfire2003}
{Wolfire}, M.~G., {McKee}, C.~F., {Hollenbach}, D., \& {Tielens}, A.~G.~G.~M.
  2003, \apj, 587, 278

\bibitem[{{Zhang} {et~al.}(2014){Zhang}, {Moscadelli}, {Sato}, {Reid},
  {Menten}, {Zheng}, {Brunthaler}, {Dame}, {Xu}, \& {Immer}}]{Zhang2014}
{Zhang}, B., {Moscadelli}, L., {Sato}, M., {et~al.} 2014, \apj, 781, 89

\end{thebibliography}

\end{document}